\documentclass[12pt]{article}
\usepackage{amsmath,amssymb,theorem,cite,epsfig,url,psfrag,eepic,mathtools}
\usepackage[matrix,arrow,curve]{xy}
\usepackage{indentfirst}
\advance \topmargin by -\headheight
\advance \topmargin by -\headsep
\evensidemargin \oddsidemargin
\def\Maketitle{{\def\newpage{}\maketitle}}
\makeatletter
\def\Appendix{\appendix
  \def\@seccntformat##1{Appendix~\csname the##1\endcsname.~~}}
\makeatother
\makeatletter
\@addtoreset{equation}{section}

\makeatother
\theorembodyfont{\sffamily}

\def\XXint#1#2#3{{\setbox0=\hbox{$#1{#2#3}{\int}$}
\vcenter{\hbox{$#2#3$}}\kern-.5\wd0}}

\def\ñth{{\textrm{\,ñth}}}
\begin{document}
\title{\textbf{Parafermionic Liouville field theory \\ and instantons on  ALE spaces}\vspace*{1cm}}
\author{M.~N.~Alfimov$^{1,2}$\footnote{E-mail: \; alfimov-mihail@rambler.ru}  \,and G.~M.~Tarnopolsky$^{3}$\footnote{E-mail: \; hetzif@itp.ac.ru}\vspace*{10pt}\\[\medskipamount]
$^1$~\parbox[t]{0.88\textwidth}{\normalsize\it\raggedright
P.N. Lebedev Physical Institute, Leninskiy pr. 53, 119991 Moscow, Russia}\vspace*{5pt}\\[\medskipamount]
$^2$~\parbox[t]{0.88\textwidth}{\normalsize\it\raggedright
Moscow Institute of Physics and Technology, Institutskiy per. 9, 141700 Dolgoprudny, Russia}\vspace*{5pt}\\[\medskipamount]
$^3$~\parbox[t]{0.88\textwidth}{\normalsize\it\raggedright
Landau Institute for Theoretical Physics,
142432 Chernogolovka, Russia}\vspace*{5pt}}

\date{}

\rightline{\texttt{\today}}

\Maketitle
\begin{abstract}\vspace*{10pt}
In this paper we study the correspondence between the $\hat{\textrm{su}}(n)_{k}\oplus \hat{\textrm{su}}(n)_{p}/\hat{\textrm{su}}(n)_{k+p}$ coset conformal
field theories and $\mathcal{N}=2$ $SU(n)$ gauge theories on $\mathbb{R}^{4}/\mathbb{Z}_{p}$. Namely  we check the correspondence between the $SU(2)$ Nekrasov partition function on
$\mathbb{R}^{4}/\mathbb{Z}_{4}$ and the conformal blocks of the $S_{3}$ parafermion algebra (in $S$ and $D$ modules).
We find that they are equal up to the $U(1)$-factor as it was in all cases of AGT-like relations.
Studying the structure of the instanton partition function on $\mathbb{R}^4/\mathbb{Z}_p$ we also find some evidence
that this correspondence with arbitrary $p$ takes place up to the $U(1)$-factor.
\end{abstract}

\tableofcontents

\section{Introduction}

Two years ago in the seminal paper \cite{Alday:2009aq} it was proposed that $\mathcal{N}=2$
supersymmetric 4d gauge theories with $SU(2)$ gauge symmetry are related to the 2d
conformal field theory. Since then it has been studied in a lot of papers from
the side of gauge theories, conformal field theory
\cite{Fateev:2009aw, Hadasz:2010xp, Poghossian:2009mk, Mironov:2009qt, Mironov:2009qn},
 matrix models \cite{Dijkgraaf:2009pc, Cheng:2010yw, Mironov:2010pi, Mironov:2010qe}
 and abstract math \cite{Braverman:2010ef, Awata:2011fk}.

The works \cite{Belavin:2011pp, Nishioka:2011jk} proposed the
connection between $SU(n)$ gauge theories on
$\mathbb{R}^4/\mathbb{Z}_p$ and coset conformal field theories based
on the coset $\hat{\textrm{su}}(n)_{k}\oplus
\hat{\textrm{su}}(n)_{p}/\hat{\textrm{su}}(n)_{k+p}$, where $p$ is a
positive integer and $k$ is a free parameter. The first non-trivial
checks for this relation for $p=2$ were made in
\cite{Bonelli:2011jx, Belavin:2011tb, Bonelli:2011kv} and for $p=4$
in the recent paper by N.Wyllard \cite{Wyllard:2011mn}. The explicit
calculations for the case $(n,p)=(2,4)$ were done there in some
particular modules and Whittaker  limit.

In our paper we proceed with the study of the case $(n,p)=(2,4)$.
Namely, we propose the explicit equality between particular Nekrasov
instanton partition function on $\mathbb{R}^4/\mathbb{Z}_4$ and
$S_3$ parafermion four-point conformal blocks in the so called $S$
and $D$ modules\cite{Fateev:1985ig}. We have checked the
correspondence  for the first four levels. We show that this
equality holds up to the so called $U(1)$-factor, which is the same
as in all $SU(2)$ AGT-like relations. It can be schematically
represented as

\begin{equation}
\mathcal{Z}_{\textrm{instanton}}(z)=(1-z)^{A}
\mathcal{F}_{\textrm{conformal block}}(z), \notag
\end{equation}
where $A$ depends on parameters of conformal block (or instanton
partition function). \noindent Then in the end of the paper  we
argue, based on explicit evaluation of the instanton partition
function on $\mathbb{R}^4/\mathbb{Z}_p$ for $p=2,...,7$, that such a
relation holds for all $p$. We think that these general relations
can be useful for obtaining the answers for the
 conformal blocks in the parafermion theories using the gauge theory calculations.

The paper is organized as follows. Section 2 is devoted to the
instanton counting on the space $\mathbb{R}^{4}/\mathbb{Z}_{p}$. In
Section 3 we calculate the Nekrasov partition function on
$\mathbb{R}_4/\mathbb{Z}_4$. In Section 4 we remind some facts about
the extended symmetry algebra of $p=4$ parafermions and derive the
expressions for the conformal blocks of this algebra. In Section 5
we relate instanton partition function to the parafermion conformal
blocks. This relation is the main result of our paper. In Appendices
A and B we derive the commutation relations for the vertex operators
in $S$ and $D$ modules respectively. In Appendices C and D we write
out the Gram/Shapovalov matrices and matrix elements for the levels
7/4 and 2 respectively.
\section{Instanton counting on $\mathbb{R}^{4}/\mathbb{Z}_{p}$}

In this section we briefly review the  counting of the instanton partition function for
$\mathcal{N}=2$ $SU(n)$ gauge theory on $\mathbb{R}^{4}/\mathbb{Z}_{p}$ \cite{Fucito:2004ry, Fucito:2006kn}. First we describe the case of the pure $\mathcal{N}=2$ $SU(n)$ gauge theory and
then propose more general formulas.

\begin{figure}[h!]
  \begin{center}
    \includegraphics[width=12cm]{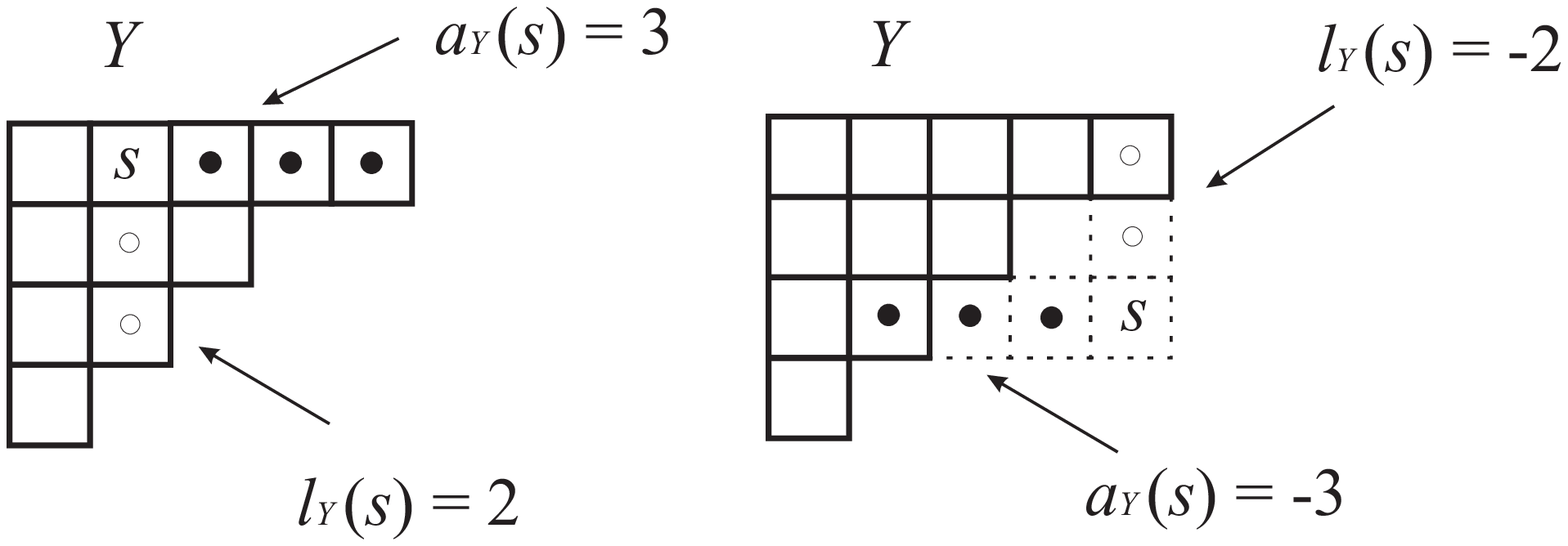}\\
  \end{center}
  \caption{The leg $l_{Y}(s)$ and the arm $a_{Y}(s)$ of the Young diagrams.}\label{examp}
\end{figure}

 The instanton partition function for the
pure $\mathcal{N}=2$ $SU(n)$ gauge theory on $\mathbb{R}^{4}$  reads
\cite{Flume:2002az}
\begin{align}
\mathcal{Z}^{(n)}(\vec{P}|z) =
\sum_{k=0}^{\infty}z^{k}\sum_{\substack{\vec{Y} \\|\vec{Y}|=k}}
\prod_{i,j=1}^{n} \prod_{s\in Y_{i}}
\frac{1}{E_{Y_{i},Y_{j}}(s|P_{i}-P_{j})(Q-E_{Y_{i},Y_{j}}(s|P_{i}-P_{j}))},
\end{align}
where the sum is over all sets of $n$ Young diagrams
$\vec{Y}=(Y_{1},...,Y_{n})$, $k = |\vec{Y}|$ is the total number of
boxes in $\vec{Y}$, $\vec{P}=(P_{1},...,P_{n})$ is the vev of the
adjoint scalar, $s$ denotes a box in the Young diagram $Y_{i}$, and
\begin{align}
E_{Y,W}(P|s)= P - l_{W}(s)b^{-1}+(a_{Y}(s)+1)b,
\end{align}
where $a_{Y}(s)$ and $l_{Y}(s)$ is the arm and the leg length respectively, i.e. the number of boxes in $Y$ to the right  and below the box $s\in Y$, see  the figure \ref{examp}.

\noindent In order to compare the answers from both gauge and conformal
field theories and for further convenience we have already
passed from the deformation parameters $\epsilon_{1}, \epsilon_{2}$
\cite{Nekrasov:2002qd, Moore:1997dj}  to the parameter $b$,
which parameterize the central charge $c$ in conformal field theory:
\begin{align}
\epsilon_{1} = b^{-1}, \quad \epsilon_{2} = b
\end{align}
and $Q= \epsilon_{1}+\epsilon_{2}= b+b^{-1}$.

\begin{figure}[h!]
  \begin{center}
    \includegraphics[width=3.0cm]{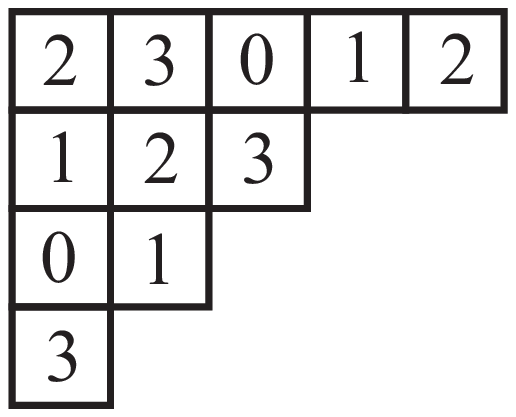}\\
  \end{center}
  \caption{The colored Young diagram $Y=\{5,3,2,1\}$ with the charge $q=2$ and for the case of $p=4$.}\label{exam}
\end{figure}

In the case of instantons on $\mathbb{R}^{4}/\mathbb{Z}_{p}$, we
have the similar structure of the partition function, but with some
differences, which we are going to describe now. One ascribes a
$\mathbb{Z}_{p}$ charge $q_{i}$, $i =1,...,n$ to each Young diagram
$Y_{i}$, where $q_{i}$ can take values $0,1,...,p-1$.  We denote
this as $Y_{i}^{q_{i}}$. It is convenient to color the Young
diagrams in $p$ colors as follows: the box with coordinates $(i,j)$
of the Young diagram $Y^{q}$ has $r \equiv q+i-j \; \textrm{mod} \;
p $ color, see for example the figure \ref{exam}.

Then we introduce two  $p$-dimensional vectors $\{n_{r}\}$ and
$\{k_{r}\}$, where $r=0,1,...,p-1$.  The integer $n_{r}$ is the
number of Young diagrams with the charge $q=r$ and $k_{r}$ is the
number of boxes with color $r$ in all Young diagrams
($Y_{1}^{q_{1}},...,Y_{n}^{q_{n}}$). Thus $\sum_{r}n_{r}=n$ and
$\sum_{r}k_{r}=k$.

Vectors $\{n_{r}\}$ and $\{k_{r}\}$ are related with topological
characteristics of instantons:
\begin{align} c_{1}(E)=
\sum_{r=0}^{p-1}(n_{r}-2k_{r}+k_{r+1}+k_{r-1})c_{1}(T_{r}),
\label{rel1}
\end{align}
where $c_{1}(E)$ is the first Chern class of gauge bundle $E$ and
$c_{1}(T_{r})$ is the first Chern class of vector bundle $T_{r}$ on
ALE space. The first Chern class $c_{1}(T_{r})$ of the $T_{r}$
bundles, $r\neq 0$ ($c_{1}(T_{0})=0$), forms a basis of the second
cohomology group. In this paper we only consider the case
$c_{1}(E)=0$. Therefore we obtain the equations
\begin{align}
0=n_{r}-2k_{r}+k_{r+1}+k_{r-1}, \quad \textrm{for} \quad r=1,...,p-1. \label{rel2}
\end{align}
For simplicity  we make the shift\footnote{One should notice that
$\delta k_{r}$ is the difference between the number of boxes with
the color $r$ and with the color $0$ in all Young diagrams.}
$k_{r}=k_{0}+\delta k_{r}$, then we have for $\delta k_{r}$
\begin{align}
\delta k_{r}=\sum_{l=1}^{p-1}C_{rl}n_{l}, \quad \textrm{where} \quad C_{rl}
= \textrm{min}(r,l)- \frac{rl}{p} . \label{deltak}
\end{align}
Below we will consider only the  case of $n=2$. It is clear that
$n_{r}$ and $k_{r}$ are integers. Due to
 this constraints there are following variants
 of vector $\{n_{r}\}$: the first variant: $n_{0}=2$, $n_{r}=0, r>0$; the second:
 $n_{r}=n_{p-r}=1$
 for $r=1,...,\lfloor\frac{p}{2} \rfloor$; and the third $n_{p/2}=2$ in case of  $p$  even.
 Therefore for the first few values of $p$ we have
the following series:
\begin{align}
\begin{tabular}{|c|c|c||c|c|}
  \hline
  $p=2$ & $\{n_{r}\}$ & $\{\delta k_{r}\}$ & $q_{1}$ & $q_{2}$ \\
  \hline
  1. & $(2,0)$ & $(0,0)$ & 0 & 0 \\
  \hline
  2. & $(0,2)$ & $(1,1)$ & 1 & 1 \\
  \hline
  \hline
  $p=3$ & $\{n_{r}\}$ & $\{\delta k_{r}\}$ & $q_{1}$ & $q_{2}$ \\
  \hline
  1. & $(2,0,0)$ & $(0,0,0)$ & 0 & 0 \\
  \hline
  2. & $(0,1,1)$ & $(0,1,1)$ & 1 & 2 \\
  \hline
  \hline
  $p=4$ & $\{n_{r}\}$ & $\{\delta k_{r}\}$ & $q_{1}$ & $q_{2}$ \\
  \hline
  1. & $(2,0,0,0)$ & $(0,0,0,0)$ & 0 & 0 \\
  \hline
  2. & $(0,1,0,1)$ & $(0,1,1,1)$ & 1 & 3 \\
  \hline
  3. & $(0,0,2,0)$ & $(0,1,2,1)$ & 2 & 2 \\
  \hline
\end{tabular} \label{tabular}
\end{align}
We use the vector symbol $\vec{P}$ to stand for pairs
$\vec{P}=(P,-P)$ and  $\vec{P}'=(P',-P')$ and also $\vec{Y^{q}}=
(Y_{1}^{q_{1}},Y_{2}^{q_{2}})$ and  $\vec{W^{u}}=
(W_{1}^{u_{1}},W_{2}^{u_{2}})$. Now we can define the contribution
of the bifundamental multiplet for $SU(2)$ gauge theory on
$\mathbb{R}_4/\mathbb{Z}_p$ ($p=1$ case see for instance in
\cite{Alday:2009aq}):
\begin{align}
Z_{\textrm{bif}}^{(p)}(\alpha|\vec{P}',\vec{W^{u}},\vec{P},\vec{Y^{q}})
= \prod_{i,j=1}^{2} \prod_{s\in Y_{i}^{q_{i}\lozenge}}
(Q-E_{Y_{i},W_{j}}(P_{i}-P'_{j}|s)-\alpha)\prod_{t\in
W_{j}^{u_{j}\lozenge}}(E_{W_{j},Y_{i}}(P'_{j}-P_{i}|t)-\alpha),
\label{Zbif}
\end{align}
where
$$
E_{Y,W}(P|s)= P- l_{W}(s)b^{-1} +(a_{Y}(s)+1)b
$$
and the product is over  $s \in Y_{i}^{q_{i}}$
and $t \in W_{j}^{u_{j}}$ is restricted to the set $\lozenge$ that
includes all $s, t$ which satisfy
\begin{align}
&s \in  Y_{i}^{q_{i}\lozenge}:\;l_{W_{j}}(s)+a_{Y_{i}}(s)+1 \equiv u_{j}-q_{i} \; \textrm{mod} \; p, \notag \\
&t \in  W_{j}^{u_{j}\lozenge}:\;l_{Y_{i}}(t)+a_{W_{j}}(t)+1 \equiv q_{i}-u_{j} \; \textrm{mod} \; p,
\end{align}
(here we also have passed from gauge theory notations to the
conformal theory ones, and denote the  hypermultiplet mass by $
\alpha$.) The partition function for the theory with two fundamental
and two antifundamental matter hypermultiplets, which one should
compare with the four-point conformal blocks, reads
\begin{align}
\mathcal{Z}^{(p)}(P_{1},\alpha_{1},\alpha_{2},P_{2}|P|z)=
\sum_{\{\vec{Y}^{\vec{q}}\}}
\frac{Z_{\textrm{bif}}^{(p)}(\alpha_{1}|\vec{P}_{1},({\O}^{0},{\O}^{0}),\vec{P},
\vec{Y^{q}})
Z_{\textrm{bif}}^{(p)}(\alpha_{2}|\vec{P},\vec{Y^{q}},\vec{P}_{2},
({\O}^{0},{\O}^{0}))}
{Z_{\textrm{bif}}^{(p)}(0|\vec{P},\vec{Y^{q}},\vec{P},
\vec{Y^{q}})}\cdot z^{\frac{|\vec{Y^{q}}|}{p}}, \label{Z4}
\end{align}
where the sum goes over all the pairs of Young diagrams $\{\vec{Y^{q}}\}$ from the special series, which satisfy to the relation (\ref{rel2}).

In the next section we perform explicit calculations of the partition function (\ref{Z4}) in case of
$p=4$. As we can see from the table (\ref{tabular}) there are three series in this case. The first ($k=\sum_{r}k_{r}=|\vec{Y^{q}}|=4k_{0}$) and the third ($k=4k_{0}+4$) series make the contribution to the integer powers of parameter $z$, whereas the second series ($k=4k_{0}+3$)
gives $z^{k_{0}+3/4}$. Thus one can rewrite (\ref{Z4}) for $p=4$ as
\begin{align}
\mathcal{Z}^{(4)}(P_{1},\alpha_{1},\alpha_{2},P_{2}|P|z)=
\sum_{i=1}^{3}\mathcal{Z}^{(4,i\,\textrm{series})}(P_{1},\alpha_{1},\alpha_{2},P_{2}|P|z), \label{series0}
\end{align}
where
\begin{align}
\mathcal{Z}^{(4,i\,\textrm{series})}(P_{1},\alpha_{1},\alpha_{2},P_{2}|P|z)
= \sum_{\{\vec{Y}^{\vec{q}}\}\in i\, \textrm{series}}
\frac{Z_{\textrm{bif}}^{(4)}(\alpha_{1}|\vec{P}_{1},({\O}^{0},{\O}^{0}),\vec{P},
\vec{Y^{q}})
Z_{\textrm{bif}}^{(4)}(\alpha_{2}|\vec{P},\vec{Y^{q}},\vec{P}_{2},
({\O}^{0},{\O}^{0}))}
{Z_{\textrm{bif}}^{(4)}(0|\vec{P},\vec{Y^{q}},\vec{P},
\vec{Y^{q}})}\cdot z^{\frac{|\vec{Y^{q}}|}{p}},  \label{Z4ser}
\end{align}
where $i=1,2,3$.
We can also rewrite it as
\begin{align}
\mathcal{Z}^{(4)}(P_{1},\alpha_{1},\alpha_{2},P_{2}|P|z)= \sum_{k_{0}=0}^{\infty}\mathcal{Z}_{k_{0}}^{(4,1\textrm{st}\,\textrm{series})}z^{k_{0}}
+\sum_{k_{0}=0}^{\infty}\mathcal{Z}_{k_{0}+3/4}^{(4,2\textrm{nd}\,\textrm{series})}z^{k_{0}+3/4}
+\sum_{k_{0}=0}^{\infty}\mathcal{Z}_{k_{0}+1}^{(4,3\textrm{rd}\,\textrm{series})}z^{k_{0}+1}. \label{series}
\end{align}
We  check that each of these series  coincide with the particular
 type of conformal blocks of $S_{3}$-parafermion algebra up to the $U(1)$-factor which is equal to $(1-z)^{A_{4}}$, where $A_{4}=\frac{1}{2}\alpha_{1}(Q-\alpha_{2})$ in case of $p=4$. As we will argue in Section 6 in the case of arbitrary $p$  it should has the same structure with $A_{p}=\frac{2}{p}\alpha_{1}(Q-\alpha_{2})$.

\section{Instanton contributions corresponding
to the $S$ and $D$ \\ modules of the $S_{3}$ parafermion algebra}
In this section we make the explicit calculations of  the partition function
 $\mathcal{Z}^{(p)}(P_{1},\alpha_{1},\alpha_{2},P_{2}|P|z)$ for the case of
 $p=4$ up to the level $z^{2}$, therefore
we need to consider all Young diagrams which contribute to the partition
function at the levels $z^{3/4}, z, z^{7/4}$ and $z^{2}$.
For simplicity of the answers we will use the notations from the conformal
field theory side, such as
\begin{align}
\Delta_{P_{i}}= \frac{1}{4}(\frac{Q^{2}}{4}-P_{i}^{2}), \quad
 \Delta_{\alpha_{i}}=\frac{1}{4}\alpha_{i}(Q-\alpha_{i}), \quad c = 2+\frac{3}{2}Q^{2}, \quad Q=b+b^{-1}.
\end{align}

\begin{figure}[h!]
  \begin{center}
    \includegraphics[width=12.5cm]{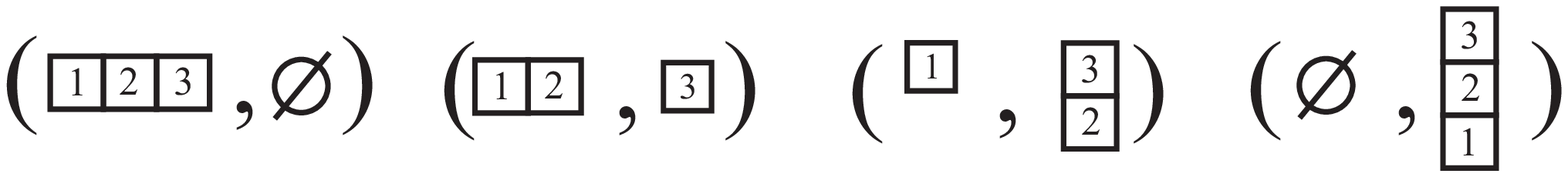}\\
  \end{center}
  \caption{The Young diagrams from the 2nd series at the level  $z^{3/4}$.}\label{diag34}
\end{figure}

\noindent $\bullet\;$ The pairs of the Young diagrams which
contribute to the partition function at the level $z^{3/4}$ are $
(\{3\},\{{\O}\})$,  $(\{2\},\{1\})$, $(\{1\},\{1,1\})$,
$(\{{\O}\},\{1,1,1\})$, which are from the second series with the
charges $ q_{1}=1,\; q_{2}=3$. They are  explicitly shown in the
figure \ref{diag34}.

\noindent Using the formulas  (\ref{Z4ser}) and $(\ref{Zbif})$ for $p=4$ we have
\begin{align}
\mathcal{Z}_{3/4}^{(4,2\textrm{nd}\,\textrm{series})} = \frac{4}{(Q+2b^{-1}-2P)(Q+2b+2P)}. \label{Z34}
\end{align}
Notice that there is no symmetry between the pairs of the Young diagrams  $(Y_{1}^{q_{1}},Y_{2}^{q_{2}}) \leftrightarrow (Y_{2}^{q_{1}},Y_{1}^{q_{2}})$
due to the presence of different charges: $q_{1}\neq q_{2}$, thus  $\mathcal{Z}^{(4,2\textrm{nd}\,\textrm{series})}(P_{1},\alpha_{1},\alpha_{2},P_{2}|P|z)$ doesn't obey $P \to -P$ symmetry. In other words, the following general identity holds
\begin{align}
\mathcal{Z}^{(p, i\,\textrm{series})}_{q_{1},q_{2}}(P_{1},\alpha_{1},\alpha_{2},P_{2}|P|z) =
\mathcal{Z}^{(p, i\,\textrm{series})}_{q_{2},q_{1}}(P_{1},\alpha_{1},\alpha_{2},P_{2}|-P|z)
\end{align}
where $q_{1}$ and $q_{2}$ are the charges of the Young diagrams of
$i\,$series. This structure has the clear meaning from the conformal
field theory (CFT) point of view.

\begin{figure}[h!]
  \begin{center}
    \includegraphics[width=13.5cm]{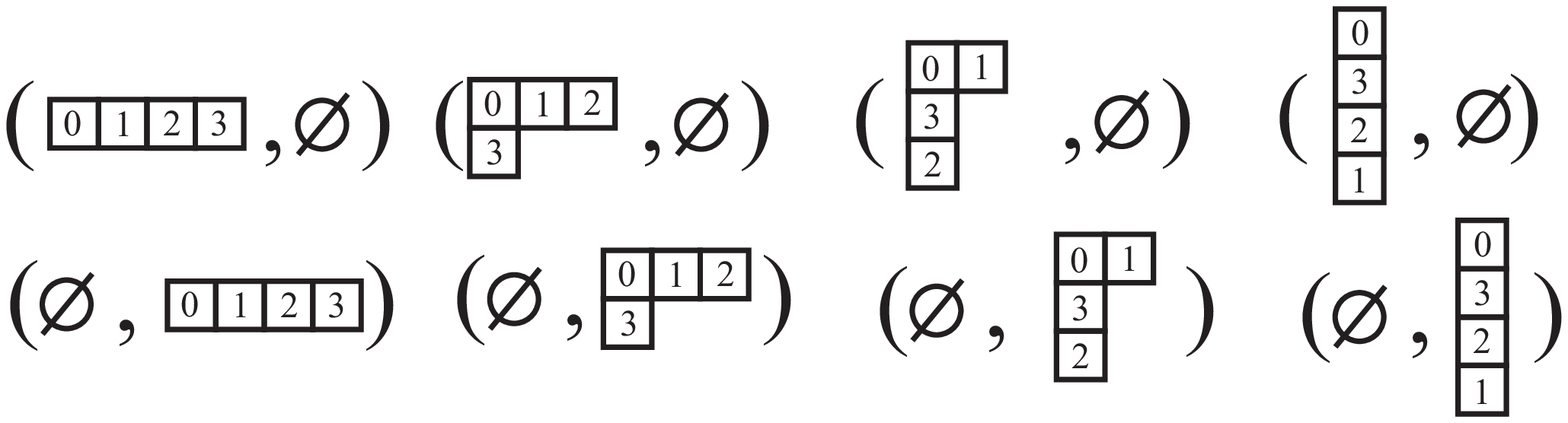}\\
  \end{center}
  \caption{The Young diagrams from  the 1st series for the level $z$.}\label{diag11}
\end{figure}

\noindent $\bullet\;$ The pairs of the Young diagrams which  contribute to the partition function at the level $z$ are divided
into the first and the third series. For the first series we have the following pairs of the Young diagrams:
$(\{4\},\{{\O}\})$, $(\{3,1\},\{{\O}\})$,  $(\{2,1,1\},\{{\O}\})$, $(\{1,1,1,1\},\{{\O}\})$,
$(\{{\O}\},\{4\})$, $(\{{\O}\},\{3,1\})$, $(\{{\O}\},\{2,1,1\})$, $(\{{\O}\},\{1,1,1,1\})$ with the charges $q_{1}=q_{2}=0$. They are explicitly shown in the figure \ref{diag11}.

\begin{figure}[h!]
  \begin{center}
    \includegraphics[width=11cm]{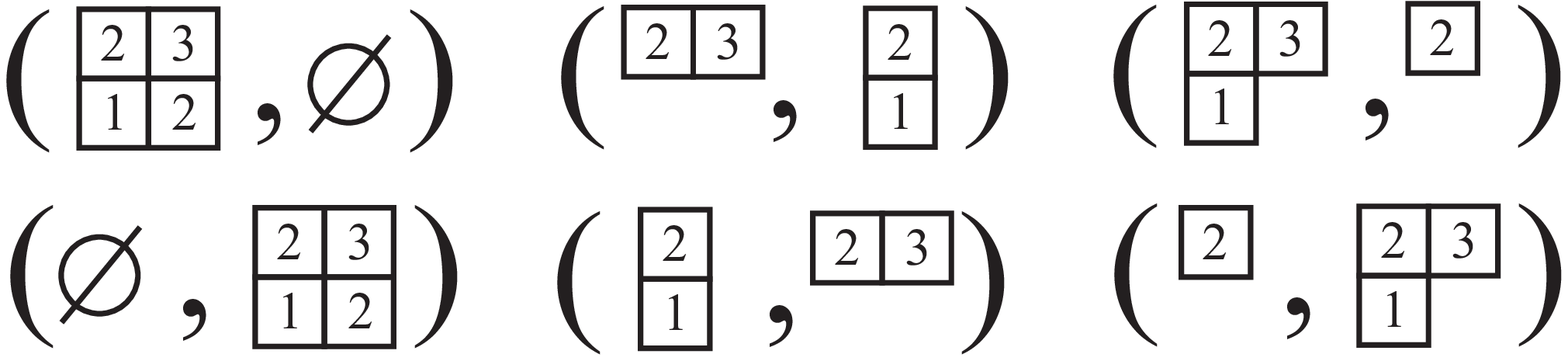}\\
  \end{center}
  \caption{The Young diagrams from the 3rd series for the level $z$.}\label{diag13}
\end{figure}

\noindent And we have for  $\mathcal{Z}_{1}^{(4,1\textrm{st}\,\textrm{series})}$
the following answer
\begin{align}
\mathcal{Z}_{1}^{(4,1\textrm{st}\,\textrm{series})} =  \frac{(\Delta_{P}-\Delta_{P_{1}}+\Delta_{\alpha_{1}})
(\Delta_{P}-\Delta_{P_{2}}+\Delta_{\alpha_{2}})}{2\Delta_{P}}-\frac{1}{2}\alpha_{1}(Q-\alpha_{2}). \label{Z11}
\end{align}
(we also imply that $\mathcal{Z}_{0}^{(4,1\textrm{st}\,\textrm{series})}=1$, and there is only one pair of the Young diagrams   $(\{{\O}^{0}\},\{{\O}^{0}\})$ for this case). The pairs of the Young diagrams for the third series are $(\{2,2\},\{{\O}\})$, $(\{2\},\{1,1\})$, $(\{2,1\},\{1\})$,
$(\{{\O}\},\{2,2\})$, $(\{1\},\{2,1\})$,  $(\{1,1\},\{2\})$, with the charges $q_{1}=q_{2}=2$. These diagrams are shown
in the figure \ref{diag13}.

\noindent Then one can obtain
\begin{align}
\mathcal{Z}_{1}^{(4,3\textrm{rd}\,\textrm{series})} =  \frac{3}{128\Delta_{P}(\Delta_{P}+\frac{c}{8}-\frac{1}{4})}. \label{Z31}
\end{align}

\begin{figure}[h!]
  \begin{center}
    \includegraphics[width=18cm]{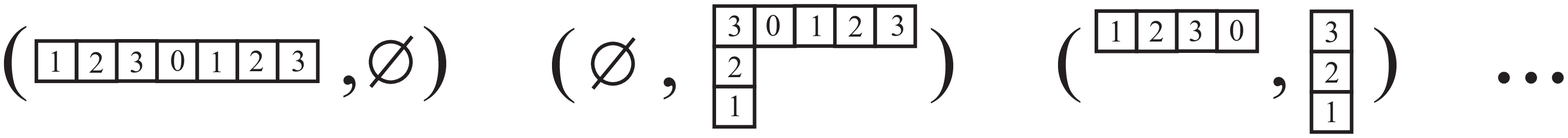}\\
  \end{center}
  \caption{The Young diagrams from the 2nd series for the level $z^{7/4}.$}\label{diag74}
\end{figure}

\noindent $\bullet\;$ There are $32$  pairs of the Young diagrams
which contribute to the partition function at the level $z^{7/4}$.
The first few such diagrams are shown in the figure \ref{diag74}.

\noindent The answer for $\mathcal{Z}_{7/4}^{(4,2\textrm{nd}\,\textrm{series})}$ is too cumbersome and we don't show it in this paper explicitly.

\noindent $\bullet\;$ For the level $z^{2}$ the $1$st series and the $3$rd series give contribution. There are $56$ pairs of the Young diagrams from the $1$st series and $48$ pairs of the Young diagrams from the $3$rd series on this level. The explicit answers for
$\mathcal{Z}_{2}^{(4,1\textrm{st}\,\textrm{series})}$ and for  $\mathcal{Z}_{2}^{(4,3\textrm{rd}\,\textrm{series})}$ are also omitted due to their enormous sizes.

\section{Conformal blocks of the $S_{3}$ parafermion algebra}

\subsection{The extended symmetry algebra of $p=4$ parafermions}

It is known that the extended symmetry algebra of the so called
$S_3$ parafermions is the spin 4/3 algebra. One can find a detailed
discussion of this algebra in \cite{Argyres:1993hz, Fateev:1985ig}.
Here we point out only the aspects relevant for our discussion.
First we write down the Operator Product Expansions (OPEs) of the
stress tensor of this algebra $T(z)$ and the fractional spin 4/3
currents $G^{\pm}(z)$

\begin{align}
    & T(z)T(w)=\frac{c}{2(z-w)^4}+\frac{2}{(z-w)^2}T(w)+\frac{1}{z-w}\partial T(w)+\cdots, \label{opett} \\
    & T(z)G^{\pm}(w)=\frac{4}{3(z-w)^2}G^{\pm}(w)+\frac{1}{z-w}\partial G^{\pm}(w)+\cdots, \label{opetg} \\
    & G^{\pm}(z) G^{\pm}(w)=\frac{\lambda^{\pm}}{(z-w)^{\frac{4}{3}}} G^{\mp}(w)+\frac{\lambda^{\mp}}{(z-w)^{\frac{1}{3}}} \partial G^{\mp}(w)+\cdots, \label{frac_ope1} \\
    & G^{\pm}(z) G^{\mp}(w)=\frac{3c}{8(z-w)^{\frac{8}{3}}}+\frac{1}{(z-w)^{\frac{2}{3}}}T(w)+\cdots, \label{frac_ope2}
\end{align}

\noindent where

\begin{equation}\label{}
    c=2+\frac{3}{2}Q^2, \qquad \lambda^{\pm}=\pm \sqrt{\frac{c-8}{6}}, \qquad Q = b+b^{-1}
\end{equation}

\noindent and dots stand for non-singular terms. The so called $\mathbb{Z}_3$-charge is associated to each field in the above algebra. One ascribes the $\mathbb{Z}_3$-charge $q=0$ to the stress tensor $T(z)$ and the $\mathbb{Z}_3$-charges $q=\pm 1$ to the fractional currents $G^{\pm}(z)$.

The next important thing to point out is the moding of the fractional currents. These modes can be determined only acting on the state $\chi_q (0)$ with the appropriate $\mathbb{Z}_3$-charge $q$

\begin{equation}\label{moding}
    G^{\pm}_{k-\frac{1 \mp q}{3}} \chi_q (0) \stackrel{\textrm{def}}{=} \oint_{\gamma} \frac{dz}{2\pi i} z^{k \pm \frac{q}{3}} G^{\pm}(z) \chi_q (0),
\end{equation}

\noindent where $\gamma$ is a contour encircling the origin and $k \in \mathbb{Z}$. From the formula (\ref{moding}) one can extract the rule that says which modes of $G^{\pm}$ can act on the state with the $\mathbb{Z}_3$-charge $q$:

\begin{equation}\label{rule_G}
    G^{\pm}_{k-\frac{1 \mp q}{3}} \left| q \right\rangle, \qquad k \in \mathbb{Z},\;\; q=0,\pm1.
\end{equation}

The commutation relations for $T(z)$ and $G^{\pm}(z)$ modes can
be easily obtained from the corresponding OPEs (\ref{opett}) and (\ref{opetg})

\begin{align}\label{}
    & \left[ L_{m}, L_{n} \right]=(m-n)L_{m+n}+\frac{c}{12}\left(m^3-m \right)\delta_{m+n,0}, \\
    & \left[ L_m, G^{\pm}_r \right]=\left(\frac{m}{3}-r \right)G^{\pm}_{m+r},
\end{align}

\noindent where $m,n \in \mathbb{Z}$ and  $r=k+\frac{1 \mp q}{3}$
with $k \in \mathbb{Z}$ and $q$ the $\mathbb{Z}_3$-charge of the
state on which the corresponding $G^{\pm}_r$ acts. It appears that
due to the fact that the OPEs (\ref{frac_ope1}) and
(\ref{frac_ope2}) of the fractional spin algebra contain the
fractional powers, it is not possible to obtain the ordinary
commutation relations but it is possible to write down the so called
generalized commutation relations
\begin{align}\label{}
    & \sum^{+\infty}_{k=0} C^{\left(-\frac{2}{3} \right)}_{k} \left( G^{+}_{\frac{q}{3}+n-k}G^{+}_{\frac{2+q}{3}+m-k}-G^{+}_{\frac{q}{3}+m-k}G^{+}_{\frac{2+q}{3}+n-k} \right)\left| q \right\rangle=\frac{\lambda^{+}}{2}(n-m)G^{+}_{\frac{2+2q}{3}+n+m}\left| q \right\rangle, \notag \\
    & \sum^{+\infty}_{k=0} C^{\left(-\frac{2}{3} \right)}_{k} \left( G^{-}_{-\frac{q}{3}+n-k}G^{-}_{\frac{2-q}{3}+m-k}-G^{-}_{-\frac{q}{3}+m-k}G^{+}_{\frac{2-q}{3}+n-k} \right)\left| q \right\rangle=\frac{\lambda^{-}}{2}(n-m)G^{+}_{\frac{2-2q}{3}+n+m}\left| q \right\rangle, \\
    & \sum^{+\infty}_{k=0} C^{\left(-\frac{1}{3} \right)}_{k} \left( G^{+}_{\frac{1+q}{3}+n-k}G^{-}_{-\frac{1+q}{3}+m-k}+G^{-}_{\frac{1+q}{3}+m-k}G^{+}_{-\frac{1+q}{3}+n-k} \right)\left| q \right\rangle=\left(L_{n+m}+\frac{3c}{16}\left(n+1+\frac{q}{3} \right)\left(n+\frac{q}{3} \right) \right) \left| q \right\rangle, \notag
\end{align}

\noindent where $m,n\in \mathbb{Z}$ and

\begin{equation}\label{}
    C^{(\nu)}_{k}=(-1)^k \left(
                           \begin{array}{c}
                             \nu \\
                             k \\
                           \end{array}
                         \right)=\frac{(-1)^k}{k!} \prod_{i=1}^{k} \left(\nu-i+1 \right).
\end{equation}

The structure of the highest weight states in this algebra is the following. There are three different modules denoted by $S$,$D$ and $R$. The highest weight states in these modules are annihilated by the generators $L_{m}$ and $G^{\pm}_r$ with $r,m>0$. The $R$-module is not relevant to our discussion. We denote the primary states as

\begin{equation}
\left| \alpha; q \right\rangle,
\end{equation}

\noindent where $\alpha$ is the Liouville parameter of the state and $q$ is the $\mathbb{Z}_3$-charge of the state. The conjugated state is denoted by $\left\langle \alpha; q \right|$. Then in these notations the highest weight state in the $S$-module is $\left| \alpha;0 \right\rangle$. In the $D$-module there are two highest weight states, i.e the highest weight state is double degenerate, and they are denoted by $\left| \alpha; \pm 1 \right\rangle$. The conformal dimensions of these highest weight states are

\begin{equation}\label{}
    \Delta^{(s)}_{\alpha}=\frac{1}{4}\alpha\left(Q-\alpha \right), \qquad \Delta^{(d)}_{\alpha}=\frac{1}{4}\alpha\left(Q-\alpha \right)+\frac{1}{12}.
\end{equation}
We use the following convention for the highest weight states in the $D$-module
\begin{equation}\label{}
    G^{\pm}_0 \left| \alpha; \pm 1 \right\rangle=\Lambda^{\pm} \left| \alpha; \mp 1 \right\rangle, \qquad \Lambda^{\pm}=\pm\sqrt{\frac{c}{24}-\Delta^{(d)}_{\alpha}}.
\end{equation}
The basis of states in each module can be chosen in the following form
\begin{equation}\label{}
    \prod_{i=1}^{u} G^{\pm}_{-r_i} \prod_{j=1}^{v} L_{-n_j} \left| \alpha;q \right\rangle.
\end{equation}

The level of the state is defined as $l=l_0+\delta$ with $l_0=\sum_{i=1}^{u}r_i+\sum_{j=1}^{v}n_j$ and where $\delta=0$ for the $S$-module and $\delta=\frac{1}{12}$ for the $D$-module. Note that in these definitions the level of the highest weight state in the $D$-module is $\frac{1}{12}$ and its dimension is $\Delta^{(d)}_{\alpha}=\Delta^{(s)}_{\alpha}+\frac{1}{12}$, which means that we count the level of the state off the level of the highest weight state in the $S$-module. Taking into account the rule (\ref{rule_G}) for acting on the states one can show that the possible values of the levels in the $S$-module are $l \in \mathbb{Z}_{\geqslant 0}$ and $l \in \mathbb{Z}_{\geqslant 0}+\frac{1}{3}$. The same in the $D$-module are $l \in \mathbb{Z}_{\geqslant 0}+\frac{1}{12}$ and $l \in \mathbb{Z}_{\geqslant 0}+\frac{3}{4}$. The hermitian conjugated generators are: $(L_n)^{\dag}=L_{-n}$, $(G^{\pm}_r)^{\dag}=-G^{\mp}_{-r}$.

\subsection{Conformal blocks}

To calculate the coefficients of the parafermionic conformal blocks we evaluate the four-point correlation function of the highest weight states in the $S$-module. We denote the vertex operator of this highest weight state as $W^{(s)}_{\alpha}(z)$. Then the object of our interest is

\begin{equation}\label{4-point function}
\langle m_1;0 | W^{(s)}_{\alpha_1}(1) W^{(s)}_{\alpha_2}(z) | m_2;0 \rangle \quad (\, \equiv \langle W^{(s)}_{m_1}(\infty) W^{(s)}_{\alpha_1}(1) W^{(s)}_{\alpha_2}(z) W^{(s)}_{m_2}(0) \rangle \; ).
\end{equation}

\noindent Here we already use the notations of the Liouville parameters like in the gauge theory to match the conformal block to the partition function. The connection of the parameters $\alpha$, $m_1$ and $m_2$ with the gauge theory parameter is
\begin{align}
\alpha=\frac{Q}{2}+P,\qquad  m_1=\frac{Q}{2}+P_1, \qquad  m_2=\frac{Q}{2}+P_2.
\end{align}
We evaluate this correlation function by inserting the complete set of states at each level
\begin{align}
\hat{\mathbf{1}}_l = \sum_{i,j}| i\rangle_{l} \times (K^{-1}_{\alpha}(l))_{ij} \times {}_{l}\langle j|, \label{set}
\end{align}
where $\{|1\rangle_{l}, |2\rangle_{l},... \}$ is the set of the descendants of primary fields at the level $l$ and $K^{-1}_\alpha (l)$ is the inverse Gram/Shapovalov matrix at the level $l$ (($K_\alpha (l))_{ij}= {}_{l}\langle i|j\rangle_{l}$). We notice that descedants states at each level depend on Liouville parameter $\alpha$ as well as the Gram/Shapovalov matrix. The corresponding conformal block is represented as the series in the fractional powers $l \in \mathbb{Z}_{\geqslant 0}$, $l \in \mathbb{Z}_{\geqslant 0}+\frac{1}{12}$, $l \in \mathbb{Z}_{\geqslant 0}+\frac{1}{3}$ and $l \in \mathbb{Z}_{\geqslant 0}+\frac{3}{4}$ of $z$. But as we already know from the gauge theory partition function (\ref{series}) that it doesn't contain the powers of the expansion parameter $l \in \mathbb{Z}_{\geqslant 0}+\frac{1}{12}$ and $l \in \mathbb{Z}_{\geqslant 0}+\frac{1}{3}$. Then we make a conclusion that to match the gauge theory result we must only consider the contributions to the correlation function (\ref{4-point function}) arising from the introduction of the complete sets (\ref{set}) of states at the levels $l \in \mathbb{Z}_{\geqslant 0}$ and $l \in \mathbb{Z}_{\geqslant 0}+\frac{3}{4}$. As we will see further there are two types of conformal blocks on the levels $l \in \mathbb{Z}_{\geqslant 0}$:
\begin{align}\label{}
    & \sum_{l=0,1,2,\ldots} \sum_{i,j} \langle m_1;0 | W^{(s)}_{\alpha_1}(1) | i\rangle_{l}\times \left( K^{-1}_{\alpha}(l) \right)_{ij}\times  {}_{l}\langle j| W^{(s)}_{\alpha_2}(z) | m_2;0 \rangle= \notag \\
    &\qquad\quad =C^{\alpha}_{m_1,\alpha_1} C^{m_2}_{\alpha,\alpha_2} \times z^{\Delta^{(s)}_{\alpha}-\Delta^{(s)}_{\alpha_2}-\Delta^{(s)}_{m_2}} \, \mathcal{F}^{(1)}(\Delta^{(s)}_{m_1},\Delta^{(s)}_{\alpha_1},\Delta^{(s)}_{\alpha_2},\Delta^{(s)}_{m_2} | \Delta^{(s)}_{\alpha} | z )+ \notag \\
    & \qquad\qquad+\tilde{C}^{\alpha}_{m_1,\alpha_1} \tilde{C}^{m_2}_{\alpha,\alpha_2} \times z^{\Delta^{(s)}_{\alpha}-\Delta^{(s)}_{\alpha_2}-\Delta^{(s)}_{m_2}} \, \mathcal{F}^{(3)}(\Delta^{(s)}_{m_1},\Delta^{(s)}_{\alpha_1},\Delta^{(s)}_{\alpha_2},\Delta^{(s)}_{m_2}| \Delta^{(s)}_{\alpha} | z ), \label{conf_block_1}
\end{align}
\noindent where

\begin{align}\label{}
    & C^{\alpha}_{m_1,\alpha_1}=\langle m_1;0 | W^{(s)}_{\alpha_1}(1)| \alpha;0 \rangle, \label{str_1} \\
    & \tilde{C}^{\alpha}_{m_1,\alpha_1}=\langle m_1;0 | \tilde{W}^{(s)}_{\alpha_1}(1)| \alpha;0 \rangle \label{str_2}
\end{align}

\noindent are the structure constants and $\tilde{W}^{(s)}_{\alpha}(z)$ is the vertex operator corresponding to the state

\begin{equation}
    | \tilde{W}^{(s)}_{\alpha} \rangle=\left( G^{+}_{-\frac{2}{3}}G^{-}_{-\frac{1}{3}}-G^{-}_{-\frac{2}{3}}G^{+}_{-\frac{1}{3}} \right) | \alpha; 0 \rangle
\end{equation}

\noindent is the descendant that is also a Virasoro primary state. We see that there are two types of conformal blocks due to the fact that there are two different types of structure constants: one for the field $W^{(s)}_{\alpha}(z)$ and the other for the field $\tilde{W}^{(s)}_{\alpha}(z)$ and these structure constants cannot be expressed through each other. We denote the conformal blocks corresponding to $W$ and $\tilde{W}$ as $\mathcal{F}^{(1)}$ and $\mathcal{F}^{(3)}$ respectively. Note that the expansion in the second conformal block starts from the first power of $z$.

The same thing can be performed at the levels $l \in \mathbb{Z}_{\geqslant 0}+\frac{3}{4}$

\begin{align}\label{}
    & \sum_{l=\frac{3}{4},\frac{7}{4},\frac{11}{4},\ldots}  \sum_{i,j}\langle m_1;0 | W^{(s)}_{\alpha_1}(1)| i\rangle_{l} \times \left( K^{-1}_{\alpha}\left(l\right) \right)_{ij}\times  {}_{l}\langle j | W^{(s)}_{\alpha_2}(z) | m_2;0 \rangle= \notag \\
    &\qquad\quad =\mathbb{C}^{(+)\alpha}_{m_1,\alpha_1} \mathbb{C}^{(+)m_2}_{\alpha,\alpha_2} \times  z^{\Delta^{(s)}_{\alpha}-\Delta^{(s)}_{\alpha_2}-\Delta^{(s)}_{m_2}}\, \mathcal{F}^{\left( 2 \right)}(\Delta^{(s)}_{m_1},\Delta^{(s)}_{\alpha_1},\Delta^{(s)}_{\alpha_2},\Delta^{(s)}_{m_2} | P | z )+ \notag \\
    &\qquad\qquad +\mathbb{C}^{(-)\alpha}_{m_1,\alpha_1} \mathbb{C}^{(-)m_2}_{\alpha,\alpha_2} \times  z^{\Delta^{(s)}_{\alpha}-\Delta^{(s)}_{\alpha_2}-\Delta^{(s)}_{m_2}}\, \mathcal{F}^{\left( 2 \right)}(\Delta^{(s)}_{m_1},\Delta^{(s)}_{\alpha_1},\Delta^{(s)}_{\alpha_2},\Delta^{(s)}_{m_2} | -P | z ) , \label{conf_block_2}
\end{align}

\noindent where $\{\left| i \right\rangle_{l}\}$ is the set of the descendants in the $D$-module at the level $l $, $K^{-1}_\alpha \left(l \right)$ is the inverse Gram/Shapovalov matrix at the level $l$ and

\begin{equation}\label{str_3}
    \mathbb{C}^{(\pm)\alpha}_{m_1,\alpha_1}=\left\langle m_1;0 \left| V^{(s)+}_{\alpha_1}(1) \right| \alpha;-1 \right\rangle \pm \left\langle m_1;0 \left| V^{(s)-}_{\alpha_1}(1) \right| \alpha;1 \right\rangle,
\end{equation}

\noindent where $V^{(s)\pm}_{\alpha}(z)$ are the vertex operators corresponding to the states

\begin{equation}
| V^{(s)\pm}_{\alpha} \rangle=G^{\pm}_{-\frac{1}{3}} | \alpha; 0 \rangle,
\end{equation}

\noindent which are the first fractional descendants of the state $| \alpha; 0 \rangle$. Note that the expansion of the conformal block $\mathcal{F}^{\left( 2 \right)}$ starts from $z^\frac{3}{4}$. Another important thing one should notice about the conformal block $\mathcal{F}^{(2)}$ is that it is not symmetric under the flip of the sign of $P$ in the parameter $\alpha=\frac{Q}{2}+P$. That's why we write $P$ instead of $\Delta^{(s)}_{\alpha}$ in the formula (\ref{conf_block_2}).

With the notations (\ref{str_1}), (\ref{str_2}) and (\ref{str_3}) we can briefly summarize the statements (\ref{conf_block_1}) and (\ref{conf_block_2})
\begin{align}\label{}
    & \sum^{+\infty}_{l=0,1,2,} \sum_{i,j} \langle m_1;0 | W^{(s)}_{\alpha_1}(1) | i \rangle_{l} \times  ( K^{-1}_{\alpha}(l))_{ij} \times {}_{l}\langle i | W^{(s)}_{\alpha_2}(z) | m_2;0\rangle= \notag \\
    & \qquad\quad=z^{\Delta^{(s)}_{\alpha}-\Delta^{(s)}_{\alpha_2}-\Delta^{(s)}_{m_2}}\left( C^{\alpha}_{m_1,\alpha_1} \cdot C^{m_2}_{\alpha, \alpha_2}  \times \mathcal{F}^{(1)}(\Delta^{(s)}_{m_1},\Delta^{(s)}_{\alpha_1},\Delta^{(s)}_{\alpha_2},\Delta^{(s)}_{m_2} | \Delta^{(s)}_{\alpha}| z )+ \right. \notag \\
    & \qquad\qquad\qquad\qquad\qquad\quad \left. +\tilde{C}^{\alpha}_{m_1,\alpha_1} \cdot \tilde{C}^{m_2}_{\alpha, \alpha_2} \times \mathcal{F}^{(3)}(\Delta^{(s)}_{m_1},\Delta^{(s)}_{\alpha_1},\Delta^{(s)}_{\alpha_2},\Delta^{(s)}_{m_2}| \Delta^{(s)}_{\alpha}| z ) \right), \label{conf_1} \\
    & \sum^{+\infty}_{l=\frac{3}{4},\frac{7}{4}\frac{11}{4},} \sum_{i,j} \langle m_1;0 | W^{(s)}_{\alpha_1}(1) | i \rangle_{l} \times \left( K^{-1}_{\alpha}\left(l \right) \right)_{ij} \times  {}_{l}\langle j| W^{(s)}_{\alpha_2}(z) | m_2;0\rangle= \notag \\
    & \qquad\quad=z^{\Delta^{(s)}_{\alpha}-\Delta^{(s)}_{\alpha_2}-\Delta^{(s)}_{m_2}} \left( \mathbb{C}^{(+)\alpha}_{m_1,\alpha_1} \cdot \mathbb{C}^{(+)m_2}_{\alpha, \alpha_2} \times \mathcal{F}^{\left( 2 \right)}(\Delta^{(s)}_{m_1},\Delta^{(s)}_{\alpha_1},\Delta^{(s)}_{\alpha_2},\Delta^{(s)}_{m_2} | P | z )+ \right. \notag \\
    & \qquad\qquad\qquad\qquad\qquad\quad \left. +\mathbb{C}^{(-)\alpha}_{m_1,\alpha_1} \cdot \mathbb{C}^{(-)m_2}_{\alpha, \alpha_2} \times \mathcal{F}^{\left( 2 \right)}(\Delta^{(s)}_{m_1},\Delta^{(s)}_{\alpha_1},\Delta^{(s)}_{\alpha_2},\Delta^{(s)}_{m_2} | -P | z ) \right). \label{conf_2}
\end{align}
And for the sake of convenience we admit the following
notations for the coefficients in the conformal block expansions
\begin{align}\label{}
    & \mathcal{F}^{(1)}(\Delta^{(s)}_{m_1},\Delta^{(s)}_{\alpha_1},\Delta^{(s)}_{\alpha_2},\Delta^{(s)}_{m_2} | \Delta^{(s)}_{\alpha} | z )=1+\mathcal{F}^{(1)}_1 z+\mathcal{F}^{(1)}_2 z^2+\cdots, \notag \\
    & \mathcal{F}^{(2)}(\Delta^{(s)}_{m_1},\Delta^{(s)}_{\alpha_1},\Delta^{(s)}_{\alpha_2},\Delta^{(s)}_{m_2} | P | z )=\mathcal{F}^{(2)}_{\frac{3}{4}} z^{\frac{3}{4}}+\mathcal{F}^{(2)}_{\frac{7}{4}} z^{\frac{7}{4}}+\cdots, \\
    & \mathcal{F}^{(3)}(\Delta^{(s)}_{m_1},\Delta^{(s)}_{\alpha_1},\Delta^{(s)}_{\alpha_2},\Delta^{(s)}_{m_2} | \Delta^{(s)}_{\alpha} | z )=\mathcal{F}^{(3)}_1 z+\mathcal{F}^{(3)}_2 z^2+\cdots. \notag
\end{align}

The above discussion drives us to the conclusion that two key ingredients we need to calculate are the inverse Gram/Shapovalov matrix at the particular level and the matrix elements of the states at the particular level

\begin{equation}\label{}
    \langle m_1;0 | W^{(s)}_{\alpha_1}(1) | i \rangle_{l}, \qquad {}_{l}\langle j| W^{(s)}_{\alpha_2}(z) | m_2;0 \rangle,
\end{equation}

\noindent where $| i\rangle_{l}$ is a state at the level $l$. To evaluate these matrix elements we must derive the commutation relations for the vertex operators.

The derivation of the commutation relations for $W^{(s)}_{\alpha}(z)$ and $\tilde{W}^{(s)}_{\alpha}(z)$ is presented in the Appendix \ref{comm_c}. Here we just cite the result

\begin{align}\label{}
    & [L_m, W^{(s)}_{\alpha}(z)]=z^{m}\partial W^{(s)}_{\alpha}(z)+(m+1)\Delta^{(s)}_{\alpha} W^{(s)}_{\alpha}(z), \\
    & [G^{\pm}_r, W^{(s)}_{\alpha}(z) ]=z^{r+\frac{1}{3}}V^{(s)\pm}_{\alpha}(z), \label{cr} \\
    & [L_m, V^{(s)\pm}_{\alpha}(z) ]=z^{m}\partial V^{(s)\pm}_{\alpha}(z)+(m+1)\left(\Delta^{(s)}_{\alpha}+\frac{1}{3} \right) V^{(s)\pm}_{\alpha}(z), \\
    & \sum_{l=0}^{+\infty} C^{\left(-\frac{1}{3} \right)}_l \left( z^{l} G^{\pm}_{r-l-\frac{1}{3}} V^{(s)\mp}_{\alpha}(z) + z^{-l-\frac{1}{3}} V^{(s)\mp}_{\alpha}(z) G^{\pm}_{r+l} \right)=\notag\\
    &\qquad\qquad\qquad=z^{r-\frac{2}{3}}\left(r+\frac{1}{3} \right) \Delta^{(s)}_{\alpha} W^{(s)}_{\alpha}(z)+z^{r+\frac{1}{3}}\left(\frac{1}{2}\partial W^{(s)}_{\alpha}(z) \pm \tilde{W}^{(s)}_{\alpha}(z) \right). \label{gcr_s}
\end{align}

Also we derived the commutation relations in the $D$-module in the Appendix \ref{comm_d}. These relations may be useful for those who want to calculate the matrix elements of the $D$-module fields. Here we again cite only the result

\begin{align*}\label{}
    & [L_m, W^{(d)\pm}_{\alpha}(z)]=z^{m}\partial W^{(d)\pm}_{\alpha}(z)+(m+1)\Delta^{(d)}_{\alpha} W^{(d)\pm}_{\alpha}(z), \\
    & [L_m, V^{(d)(\pm)}_{\alpha}(z)]=z^{m}\partial V^{(d)(\pm)}_{\alpha}(z)+(m+1)\left(\Delta^{(d)}_{\alpha}+\frac{2}{3} \right) V^{(d)(\pm)}_{\alpha}(z), \\
    & \sum_{l=0}^{+\infty} C^{(-\frac{2}{3})}_l \left( z^{l} G^{\pm}_{r-l-\frac{2}{3}} W^{(d)\pm}_{\alpha}(z)-z^{-l-\frac{2}{3}} W^{(d)\pm}_{\alpha}(z) G^{\pm}_{r+l} \right)= \notag \\
    & \qquad\qquad=z^{r-\frac{2}{3}}\left(r+\frac{1}{3} \right)\Lambda^{\pm} W^{(d)\mp}_{\alpha}(z)+z^{r+\frac{1}{3}}\left( \frac{2\Lambda^{\pm}}{3\Delta^{(d)}_{\alpha}}\partial W^{(d)\mp}_{\alpha}(z)+\tilde{W}^{(d)\mp}_{\alpha}(z) \right), \\
    & \sum_{l=0}^{+\infty} C^{(-\frac{1}{3})}_l \left( z^l G^{\pm}_{r-l-\frac{1}{3}} W^{(d)\mp}_{\alpha}(z) + z^{-l-\frac{1}{3}} W^{(d)\mp}_{\alpha}(z) G^{\pm}_{r+l} \right)=\frac{1}{2}z^{r+\frac{1}{3}} \left(V^{(d)(+)}_{\alpha}(z) \pm V^{(d)(-)}_{\alpha}(z) \right),
\end{align*}

\begin{align*}
     [G^{\pm}_r, V^{(d)+}_{\alpha}(z) ]=&z^{r-\frac{2}{3}}\left(r+\frac{1}{3} \right) \left(\Delta^{(d)}_{\alpha}+\frac{c}{12}+\lambda^{\pm} \Lambda^{\mp} \right) W^{(d)\pm}_{\alpha}(z)+ \notag \\
     &+z^{r+\frac{1}{3}}\frac{\Delta^{(d)}_{\alpha}+\frac{c}{12}+\lambda^{\pm} \Lambda^{\mp}}{3\Delta^{(d)}_{\alpha}} \partial W^{(d)\pm}_{\alpha}(z)-z^{r+\frac{1}{3}} \left(\Lambda^{\pm}-\frac{1}{2}\lambda^{\pm} \right) \tilde{W}^{(d)\pm}_{\alpha}(z), \\
    [G^{\pm}_r, V^{(d)-}_{\alpha}(z) ]=&z^{r-\frac{2}{3}}\left(r+\frac{1}{3} \right) \left(\Delta^{(d)}_{\alpha}+\frac{c}{12}-\lambda^{\pm} \Lambda^{\mp} \right) W^{(d)\pm}_{\alpha}(z) \mp \notag \\
    & \mp z^{r+\frac{1}{3}}\frac{\Delta^{(d)}_{\alpha}+\frac{c}{12}-\lambda^{\pm} \Lambda^{\mp}}{3\Delta^{(d)}_{\alpha}} \partial W^{(d)\pm}_{\alpha}(z) \pm z^{r+\frac{1}{3}} \left(\Lambda^{\pm}+\frac{1}{2}\lambda^{\pm} \right) \tilde{W}^{(d)\pm}_{\alpha}(z),
\end{align*}

\noindent where $V^{(d)(\pm)}_{\alpha}(z)$ and $\tilde{W}^{(d)\pm}_{\alpha}(z)$ are the vertex operators corresponding respectively to the states

\begin{align*}\label{}
    & | V^{(d)(\pm)}_{\alpha} \rangle=G^{+}_{-\frac{2}{3}} | W^{(d)-}_{\alpha} \rangle \pm G^{-}_{-\frac{2}{3}} | W^{(d)+}_{\alpha} \rangle, \\
    & | \tilde{W}^{(d)\pm}_{\alpha} \rangle=G^{\mp}_{-1} | W^{(d)\mp}_{\alpha} \rangle-\frac{2\Lambda^{\mp}}{3\Delta^{(d)}_{\alpha}} L_{-1} | W^{(d)\pm}_{\alpha} \rangle.
\end{align*}

\subsubsection{Level 3/4}
The basis of states at this level can be chosen in the following way

\begin{align}\label{}
    & |1\rangle_{3/4} = G^{+}_{-\frac{2}{3}} \left|\alpha; -1 \right\rangle, \\
    & |2\rangle_{3/4} = G^{-}_{-\frac{2}{3}} \left|\alpha; 1 \right\rangle. \notag
\end{align}
We must evaluate the following quantity
\begin{equation}\label{}
    \sum_{i,j=1}^{2} \langle m_1;0 | W^{(s)}_{\alpha_1}(1) |i \rangle_{3/4}\times (K^{-1}_{\alpha}(3/4))_{ij}\times {}_{3/4}\langle j| W^{(s)}_{\alpha_2}(z)|m_2;0 \rangle. \label{formula_cb_1}
\end{equation}
The calculation of the corresponding Gram/Shapovalov matrix gives
\begin{equation}
K_{\alpha}\left(3/4 \right)=\left(
  \begin{array}{cc}
    -\left(\Delta^{(d)}_{\alpha}+\frac{c}{12} \right) & \sqrt{\frac{c-8}{6}} \sqrt{\frac{c}{24}-\Delta^{(d)}_{\alpha}} \\
    \sqrt{\frac{c-8}{6}} \sqrt{\frac{c}{24}-\Delta^{(d)}_{\alpha}} & -\left(\Delta^{(d)}_{\alpha}+\frac{c}{12} \right) \\
  \end{array}
\right).
\end{equation}
It is easy to find the inverse Gram/Shapovalov matrix
\begin{equation}
K^{-1}_{\alpha}\left(3/4 \right)=\left(
  \begin{array}{cc}
    -\frac{2b(1+b(2\alpha+b(4+b^2+2b\alpha-2\alpha^2)))}{(b+\alpha)(2+b^2-b\alpha)(1+2b^2-b\alpha)(1+b\alpha)} & -\frac{2b|1-b^2||1+b^2-2b\alpha|}{(b+\alpha)(2+b^2-b\alpha)(1+2b^2-b\alpha)(1+b\alpha)} \\
    -\frac{2b|1-b^2||1+b^2-2b\alpha|}{(b+\alpha)(2+b^2-b\alpha)(1+2b^2-b\alpha)(1+b\alpha)} & -\frac{2b(1+b(2\alpha+b(4+b^2+2b\alpha-2\alpha^2)))}{(b+\alpha)(2+b^2-b\alpha)(1+2b^2-b\alpha)(1+b\alpha)} \\
  \end{array}
\right). \label{gram_1}
\end{equation}
Using the commutation relations (\ref{cr}) we have
\begin{align}\label{}
    &\langle m_1;0 | W_{\alpha_1}(1)|1\rangle_{3/4}=-\left( \mathbb{C}^{(+)\alpha}_{m_1, \alpha_1}+\mathbb{C}^{(-)\alpha}_{m_1, \alpha_1} \right), \label{m_el_1}\\
    & \langle m_1;0 | W_{\alpha_2}(1) |2\rangle_{3/4}=-\left( \mathbb{C}^{(+)\alpha}_{m_1, \alpha_1}-\mathbb{C}^{(-)\alpha}_{m_1, \alpha_1} \right), \notag \\
    & {}_{3/4}\langle 1| W_{\alpha_2}(z) |m_2;0 \rangle=-z^{\Delta^{(s)}_{\alpha}-\Delta^{(s)}_{\alpha_2}-\Delta^{(s)}_{m_2}
    +\frac{3}{4}} \left( \mathbb{C}^{(+)m_2}_{\alpha, \alpha_2}+\mathbb{C}^{(-)m_2}_{\alpha, \alpha_2} \right), \notag \\
    & {}_{3/4}\langle 2| W_{\alpha_2}(z) |m_2;0 \rangle=-z^{\Delta^{(s)}_{\alpha}-\Delta^{(s)}_{\alpha_2}-\Delta^{(s)}_{m_2}
    +\frac{3}{4}} \left( \mathbb{C}^{(+)m_2}_{\alpha, \alpha_2}-\mathbb{C}^{(-)m_2}_{\alpha, \alpha_2} \right). \notag
\end{align}
Note that there are modules in the non-diagonal matrix elements in the Gram/Shapovalov matrix. To evaluate the contributions to the conformal block we must choose the sign of these modules. It appears that to match the gauge theory result (\ref{Z34}) we should choose the following sign $|1-b^2||1+b^2-2b\alpha|=(b^2-1)(2b\alpha-1-b^2)=2b(b^2-1)P$. Substituting the matrix elements (\ref{m_el_1}) and the inverse Gram/Shapovalov matrix (\ref{gram_1}) into (\ref{formula_cb_1}) after some simple algebra one has the contribution to the conformal block $\mathcal{F}^{(2)}(\Delta^{(s)}_{m_1},\Delta^{(s)}_{\alpha_1},\Delta^{(s)}_{\alpha_2},\Delta^{(s)}_{m_2}|P|z)$
\begin{equation}\label{}
    \mathcal{F}^{\left(2\right)}_{3/4}=-\frac{8b}{(b+\alpha)(2+b^2-b\alpha)}=-\frac{32}{(Q+2b+P)(Q+2b^{-1}-P)}. \label{c_b_34}
\end{equation}

One should also notice that the terms proportional to
$\mathbb{C}^{(+)\alpha}_{m_1,\alpha_1}\cdot\mathbb{C}^{(-)m_2}_{\alpha,\alpha_2}$
or
$\mathbb{C}^{(-)\alpha}_{m_1,\alpha_1}\cdot\mathbb{C}^{(+)m_2}_{\alpha,\alpha_2}$
does not appear in the final result when we calculate
(\ref{formula_cb_1}). This result is in the accordance with the
formula (\ref{conf_block_2}).

The interesting fact is that we can also match the result in case of
the different choice of the sign of the non-diagonal elements of the
inverse Gram/Shapovalov matrix. If we choose the opposite sign
$|1-b^2||1+b^2-2b\alpha|=-(b^2-1)(2b\alpha-1-b^2)=-2b(b^2-1)P$, we
obtain the result (\ref{c_b_34}) with $P$ replaced by $-P$. Then if
we simultaneously exchange the $\mathbb{Z}_4$-charges $q_{1}=1$ and
$q_{2}=3$ associated to the Young diagrams in the 2nd series in the
gauge theory (\ref{Z34}), our results will again coincide. Or,
equivalently, given the particular choice of the sign in the
Gram/Shapovalov matrix the conformal block
$\mathcal{F}^{(2)}(\Delta^{(s)}_{m_1},\Delta^{(s)}_{\alpha_1},\Delta^{(s)}_{\alpha_2},\Delta^{(s)}_{m_2}|P|z)$
multiplied by
$\mathbb{C}^{(+)\alpha}_{m_1,\alpha_1}\cdot\mathbb{C}^{(+)m_2}_{\alpha,\alpha_2}$
corresponds to one choice of the charges of the Young diagrams
$\left( Y^{q_1}_1, Y^{q_2}_2 \right)$ and the
$\mathcal{F}^{(2)}(\Delta^{(s)}_{m_1},\Delta^{(s)}_{\alpha_1},\Delta^{(s)}_{\alpha_2},\Delta^{(s)}_{m_2}|-P|z)$
multiplied by
$\mathbb{C}^{(-)\alpha}_{m_1,\alpha_1}\cdot\mathbb{C}^{(-)m_2}_{\alpha,\alpha_2}$
corresponds to the Young diagrams with exchanged
$\mathbb{Z}_4$-charges $\left( Y^{q_2}_1, Y^{q_1}_2 \right)$.

So, briefly, the exchange of the $\mathbb{Z}_4$-charges of the Young
diagrams in the gauge theory corresponds to the flip of the signs in
the matrix elements with modules of the inverse Gram/Shapovalov matrix at the levels
$l \in \mathbb{Z}_{\geqslant 0}+\frac{3}{4}$ on the CFT side.
The same situation occurs at the level $7/4$.

\subsubsection{Level 1}

The basis of states at this level can be chosen in the following way

\begin{align}\label{}
    & \left|1\right\rangle_{1} = G^{-}_{-\frac{2}{3}}G^{+}_{-\frac{1}{3}} \left|\alpha;0 \right\rangle, \\
    & \left|2 \right\rangle_{1} = G^{+}_{-\frac{2}{3}}G^{-}_{-\frac{1}{3}} \left|\alpha; 0 \right\rangle. \notag
\end{align}
We must evaluate the following quantity
\begin{equation}\label{}
    \sum_{i,j=1}^{2} \langle m_1;0 | W^{(s)}_{\alpha_1}(1) |i \rangle_{1} \times (K^{-1}_{\alpha}(1))_{ij}\times   {}_{1}\langle j| W^{(s)}_{\alpha_2}(z)|m_2;0 \rangle. \label{formula_cb_2}
\end{equation}
The Gram/Shapovalov matrix at this level is
\begin{equation}\label{}
    K_{\alpha}(1)=\frac{\Delta^{(s)}_{\alpha}}{3} \left(
      \begin{array}{cc}
        2\Delta^{(s)}_{\alpha}+\frac{c}{4}+1 & -\left( 2\Delta^{(s)}_{\alpha}+\frac{c}{4}-2 \right) \\
        -\left( 2\Delta^{(s)}_{\alpha}+\frac{c}{4}-2 \right) & 2\Delta^{(s)}_{\alpha}+\frac{c}{4}+1 \\
      \end{array}
    \right).
\end{equation}
The inverse Gram/Shapovalov matrix is
\begin{equation}\label{}
    K^{-1}_{\alpha}(1)=\left(
      \begin{array}{cc}
         \frac{8\Delta^{(s)}_{\alpha}+c+4}{2\Delta^{(s)}_{\alpha}(8\Delta^{(s)}_{\alpha}+c-2)} & \frac{8\Delta^{(s)}_{\alpha}+c-8}{2\Delta^{(s)}_{\alpha}(8\Delta^{(s)}_{\alpha}+c-2)} \\
        \frac{8\Delta^{(s)}_{\alpha}+c-8}{2\Delta^{(s)}_{\alpha}(8\Delta^{(s)}_{\alpha}+c-2)} & \frac{8\Delta^{(s)}_{\alpha}+c+4}{2\Delta^{(s)}_{\alpha}(8\Delta^{(s)}_{\alpha}+c-2)} \\
      \end{array}
    \right). \label{gram_2}
\end{equation}
After that we are ready to proceed with the calculation of matrix elements using the commutation relations (\ref{cr}) and (\ref{gcr_s})
\begin{align}\label{}
    & \langle m_1;0 | W_{\alpha_1}(1) | 1\rangle_{1}=\frac{1}{2}\left( \Delta^{(s)}_{\alpha}+\Delta^{(s)}_{\alpha_1}-\Delta^{(s)}_{m_1} \right) \times C^{\alpha}_{m_1,\alpha_1}-\tilde{C}^{\alpha}_{m_1,\alpha_1}, \label{m_el_5} \\
    & \langle m_1;0 | W_{\alpha_1}(1) | 2\rangle_{1}=\frac{1}{2}\left( \Delta^{(s)}_{\alpha}+\Delta^{(s)}_{\alpha_1}-\Delta^{(s)}_{m_1} \right)\times C^{\alpha}_{m_1,\alpha_1}+\tilde{C}^{\alpha}_{m_1,\alpha_1} \notag ,\\
    & {}_{1}\langle 1 | W_{\alpha_2}(z)| m_2;0 \rangle=z^{\Delta^{(s)}_{\alpha}-\Delta^{(s)}_{\alpha_2}-\Delta^{(s)}_{m_2}+1} \times \left(\frac{1}{2}\left( \Delta^{(s)}_{\alpha}+\Delta^{(s)}_{\alpha_2}-\Delta^{(s)}_{m_2} \right) \times C^{m_2}_{\alpha, \alpha_2}-\tilde{C}^{m_2}_{\alpha, \alpha_2} \right), \notag \\
    & {}_{1}\langle 2 | W_{\alpha_2}(z)| m_2;0 \rangle=z^{\Delta^{(s)}_{\alpha}-\Delta^{(s)}_{\alpha_2}-\Delta^{(s)}_{m_2}+1} \times \left(\frac{1}{2}\left( \Delta^{(s)}_{\alpha}+\Delta^{(s)}_{\alpha_2}-\Delta^{(s)}_{m_2} \right) \times C^{m_2}_{\alpha, \alpha_2}+\tilde{C}^{m_2}_{\alpha, \alpha_2} \right). \notag
\end{align}
Substituting the matrix elements (\ref{m_el_5}) and the inverse Gram/Shapovalov matrix (\ref{gram_2}) into (\ref{formula_cb_2}) and making again some algebra we obtain the expressions for the coefficients of the conformal blocks at the level 1:

\begin{align}\label{}
    & \mathcal{F}^{(1)}_{1}=\frac{( \Delta^{(s)}_{\alpha}+\Delta^{(s)}_{\alpha_1}-\Delta^{(s)}_{m_1} ) ( \Delta^{(s)}_{\alpha}+\Delta^{(s)}_{\alpha_2}-\Delta^{(s)}_{m_2})}{2\Delta^{(s)}_{\alpha}}, \label{cb_11} \\
    & \mathcal{F}^{(3)}_{1}=\frac{3}{2\Delta^{(s)}_{\alpha}\left( \Delta^{(s)}_{\alpha}+\frac{c}{8}-\frac{1}{4} \right)}. \label{cb_31}
\end{align}
Note that there are no contributions proportional to $C^{\alpha}_{m_1,\alpha_1}\cdot\tilde{C}^{m_2}_{\alpha, \alpha_2}$ or $\tilde{C}^{\alpha}_{m_1,\alpha_1}\cdot C^{m_2}_{\alpha, \alpha_2}$. These contributions disappear from the answer due to the symmetry of the inverse Gram/Shapovalov matrix. Note that the result is in agreement with the formula (\ref{conf_1}).

\subsubsection{Level 7/4}

Because the expressions for the conformal blocks at the level 7/4 are very cumbersome, here we just write the basis of states at the level 7/4

\begin{align}\label{}
    & | 1\rangle_{7/4}=G^{+}_{-\frac{5}{3}} \left| \alpha;-1 \right\rangle, \\
    & | 2\rangle_{7/4}=G^{-}_{-\frac{5}{3}} \left| \alpha;1 \right\rangle, \notag \\
    & | 3\rangle_{7/4}=L_{-1} G^{+}_{-\frac{2}{3}} \left| \alpha;-1 \right\rangle, \notag \\
    & | 4\rangle_{7/4}=L_{-1} G^{-}_{-\frac{2}{3}} \left| \alpha;1 \right\rangle. \notag
\end{align}
We must evaluate the following quantity
\begin{equation}\label{}
    \sum_{i,j=1}^{4} \langle m_1;0 | W^{(s)}_{\alpha_1}(1) |i\rangle_{7/4} \times (K^{-1}_{\alpha}(7/4))_{ij}
     \times\, {}_{7/4}\langle j| W^{(s)}_{\alpha_2}(z)|m_2;0 \rangle.
\end{equation}
One can find the expression for the Gram/Shapovalov matrix and matrix elements in the Appendix \ref{app_c}. Using the Gram/Shapovalov matrix and the matrix elements from the Appendix \ref{app_c} one can easily obtain the expression for the conformal block. Note that there are also no contributions proportional to $\mathbb{C}^{(+)\alpha}_{m_1,\alpha_1}\mathbb{C}^{(-)m_2}_{\alpha,\alpha_2}$ or $\mathbb{C}^{(-)\alpha}_{m_1,\alpha_1}\mathbb{C}^{(+)m_2}_{\alpha,\alpha_2}$ at this level.

\subsubsection{Level 2}

The basis of states at this level can be chosen in the following way

\begin{align}\label{}
    & \left| 1 \right\rangle_{2}=L_{-1} G^{-}_{-\frac{2}{3}} G^{+}_{-\frac{1}{3}} \left| \alpha;0 \right\rangle, \\
    & \left| 2\right\rangle_{2}=L_{-1} G^{+}_{-\frac{2}{3}} G^{-}_{-\frac{1}{3}} \left| \alpha;0 \right\rangle, \notag \\
    & \left| 3\right\rangle_{2}=G^{-}_{-\frac{5}{3}} G^{+}_{-\frac{1}{3}} \left| \alpha;0 \right\rangle, \notag \\
    & \left| 4\right\rangle_{2}=G^{+}_{-\frac{5}{3}} G^{-}_{-\frac{1}{3}} \left| \alpha;0 \right\rangle, \notag \\
    & \left| 5 \right\rangle_{2}=L_{-2} \left| \alpha;0 \right\rangle. \notag
\end{align}
We must evaluate the following quantity
\begin{equation}\label{}
    \sum_{i,j=1}^{5} \langle m_1;0 | W^{(s)}_{\alpha_1}(1) |i\rangle_{2} \times (K^{-1}_{\alpha}(2))_{ij} \times {}_{2}\langle j| W^{(s)}_{\alpha_2}(z) |m_2;0 \rangle.
\end{equation}
One can find the expression for the Gram/Shapovalov matrix and and matrix elements in the Appendix \ref{app_d}. Using the Gram/Shapovalov matrix and the matrix elements from the Appendix \ref{app_d} one can easily obtain the expression for the conformal block. Note that at the level 2 the contributions with $C^{\alpha}_{m_1,\alpha_1}\cdot \tilde{C}^{m_2}_{\alpha, \alpha_2}$ or $\tilde{C}^{\alpha}_{m_1,\alpha_1} \cdot C^{m_2}_{\alpha, \alpha_2}$ disappear again.

\section{Comparing the conformal blocks in the  $S$ and $D$ modules of the $S_{3}$  parafermion algebra with the instanton partition function on $\mathbb{R}_4/\mathbb{Z}_p$}

Recalling that in our notations $\Delta_{P_{i}}=\Delta^{(s)}_{m_i}$ for $i=1,2$ because $m_i=\frac{Q}{2}+P_i$ and using the equations for the conformal block contribution and the partition function, we find, comparing the formulae
 (\ref{Z34}) and (\ref{c_b_34}), that at the level 3/4:

\begin{align}\label{}
     &\mathcal{Z}_{3/4}^{(4,2\textrm{nd}\,\textrm{series})}=-\frac{1}{8}\mathcal{F}^{\left( 2 \right)}_{3/4}. \qquad\qquad\qquad\qquad\;
\end{align}
Comparing the formulas (\ref{Z11}), (\ref{Z31}) with (\ref{cb_11}), (\ref{cb_31}) we have at the level 1:
\begin{align}\label{}
    & \mathcal{Z}_{1}^{(4,1\textrm{st}\,\textrm{series})}=\mathcal{F}^{\left( 1 \right)}_{1}-A_4,  \qquad\qquad\qquad\qquad\\
    & \mathcal{Z}_{1}^{(4,3\textrm{rd}\,\textrm{series})}=\frac{1}{64}\mathcal{F}^{\left( 3 \right)}_{1}.\qquad\qquad \qquad\qquad
\end{align}
At the level 7/4:
\begin{align}\label{}
   & \mathcal{Z}_{7/4}^{(4,2\textrm{nd}\,\textrm{series})}=
   -\frac{1}{8}\left(\mathcal{F}^{\left( 2 \right)}_{7/4}-A_4 \mathcal{F}^{\left( 2 \right)}_{3/4} \right). \qquad\quad
\end{align}
At the level 2:
\begin{align}\label{}
    & \mathcal{Z}_{2}^{(4,1\textrm{st}\,\textrm{series})}=
    \mathcal{F}^{\left( 1 \right)}_{2}-A_4 \mathcal{F}^{\left( 1 \right)}_{1}+\frac{1}{2}A_4(A_4-1), \\
    & \mathcal{Z}_{2}^{(4,3\textrm{rd}\,\textrm{series})}=\frac{1}{64}
    \left( \mathcal{F}^{\left( 3 \right)}_{1}-A_4 \mathcal{F}^{\left( 3 \right)}_{0} \right),
\end{align}
where $A_{4}$ is
\begin{align}
A_{4}= \frac{1}{2}\alpha_{1}(Q-\alpha_{2}).
\end{align}
These checks at the particular levels lead us to the natural proposition
\begin{align}
&\mathcal{Z}^{(4,1\textrm{st}\,\textrm{series})}(P_{1},\alpha_{1},\alpha_{2},P_{2}|P|z) = (1-z)^{A_{4}}\mathcal{F}^{(1)}(\Delta_{P_{1}},\Delta_{\alpha_{1}},
\Delta_{\alpha_{2}},\Delta_{P_{2}}|\Delta_{P}|z), \label{corr_1} \\
&\mathcal{Z}^{(4,2\textrm{nd}\,\textrm{series})}(P_{1},\alpha_{1},\alpha_{2},P_{2}|P|z) = -\frac{1}{2^{3}}(1-z)^{A_{4}} \mathcal{F}^{(2)} \left( \Delta_{P_{1}},\Delta_{\alpha_{1}},
\Delta_{\alpha_{2}},\Delta_{P_{2}}|P|z \right), \label{corr_2} \\
&\mathcal{Z}^{(4,3\textrm{rd}\,\textrm{series})}(P_{1},\alpha_{1},\alpha_{2},P_{2}|P|z)= \frac{1}{2^{6}}(1-z)^{A_{4}} \mathcal{F}^{(3)} \left( \Delta_{P_{1}},\Delta_{\alpha_{1}},
\Delta_{\alpha_{2}},\Delta_{P_{2}}|\Delta_{P}|z \right). \label{corr_3}
\end{align}
The formulas (\ref{corr_1}), (\ref{corr_2}) and (\ref{corr_3}) are the main result of our paper.

\section{Concluding remarks and open problems}

This paper established the explicit view of relation between $SU(2)$ Nekrasov
instanton partition function on $\mathbb{R}^4/\mathbb{Z}_4$ and four-point
conformal blocks in $S$ and $D$ modules of the $S_3$ parafermion algebra.
This relation can be schematically written as
$\mathcal{Z}_{\textrm{instanton}}(z)=(1-z)^{A_{4}} \mathcal{F}_{\textrm{conformal block}}(z)$.

One can notice that the four-point coset conformal blocks depend on the dimensions
of the primary fields in the correlation function and the parameter $P$, which denotes
the intermediate parameter in the conformal block. As usual one parameterizes the
conformal dimensions as $\Delta_i$ by the momenta $P_i$ as $\Delta_i  = \frac{1}{p}(\frac{Q^{2}}{4}- P^2_i)$.
Therefore the coset conformal blocks are the functions  of the parameters $P^2_i$ and $P$,
so $\mathcal{F}_{\textrm{conformal block}}=f(P_{i}^{2},P|z)$. On the other hand the
instanton partition function on $\mathbb{R}^{4}/\mathbb{Z}_p$ is a function of
parameters $P_i$, $P$. We made the explicit checks for the cases of $p=2,...,7$ that the combination $$
(1-z)^{-A_p}\mathcal{Z}_{\textrm{instanton}}^{(p)}(P_i,P|z)=\phi(P_{i}^2,P|z),
$$
where $\phi$ is some function, which depends on $P_{i}^{2}$, not on $P_{i}$! Thus, one can assume that the relation
$$
\mathcal{Z}_{\textrm{instanton}}^{(p)}(P_i,P|z)=(1-z)^{A_p}\mathcal{F}_{\textrm{conformal block}}(\Delta_i, P| z)
$$
holds for all $p$ ($ A_{p}= \frac{2}{p}(
Q/2+P_{2})(Q/2-P_{3})$).

In this paper we use the particular series in  the gauge theory
$(c_{1}(E)=0)$ and the particular modules ($S$ and $D$ ) in the
conformal field theory. As it was recently shown in the work
\cite{Ito:2011mw}, one can assume that for the coincidence in other
modules of CFT we should consider other series of the Young diagrams
in the gauge theory.

Also it should be mentioned that there is another formula for the instanton partition
function on $\mathbb{R}^4/\mathbb{Z}_p$ \cite{Bruzzo:2009uc, Bruzzo:2002xf, Sasaki:2006vq, Bonelli:2011kv}.
It can be schematically viewed as
$$
\mathcal{Z}_{\textrm{instanton}}^{(p)}=l_{\textrm{factor}} \cdot
\underbrace{\mathcal{Z}_{\textrm{instanton}}^{(1)}\times \mathcal{Z}_{\textrm{instanton}}^{(1)}
\times...\times \mathcal{Z}_{\textrm{instanton}}^{(1)}}_{p},
$$ where $l_{\textrm{factor}}$ is the so called blow-up factor. This structure has a very
clear explanation from the CFT point view \cite{Alba:2010qc,
Belavin:2011bb}. Factor $l_{\textrm{factor}}$ arises from the matrix
elements of highest weight vectors of $p$ pairs of commuting
Virasoro algebras, which are constructed from another algebras.The
construction of two commuting Virasoro algebras using the
$\textrm{NSR}$ algebra and free fermion field was proposed in
\cite{Lashkevich:1992sb, Crnkovic:1989ug}. Also, relations between
parafermion Liouville structure constant \cite{Bershtein:2010wz,
Pogosian:1988ar} and Liouville structure constants
$$
 C^{(p)}_{\textrm{Parafermionic Liouville} } =
 \underbrace{C^{(1)}_{\textrm{Liouville}}\times C^{(1)}_{\textrm{Liouville}}
 \times...\times C^{(1)}_{\textrm{Liouville}}}_{p}
$$
in principle can be checked using the relations between $\Upsilon$-functions \cite{Belavin:2011bb}.

\section*{Acknowledgments}
We thank Alexey Litvinov, Alexander Belavin, Mikhail Bershtein and Rubik Poghossian for very useful discussions.

Moral support by San'kin M. was extremely important for G. T., particularly while
staying in the hospital with broken leg.

The work M.A. and G.T. was supported by 2011 Dynasty foundation grant. Also the
work of G.T. was held within the framework of the Federal programs
``Scientific and Scientific-Pedagogical Personnel of Innovational Russia'' on 2009-2013
(state contracts No. P1339 and No. 02.740.11.5165)
and was supported by cooperative CNRS-RFBR  grant PICS-09-02-93064  and
by Russian Ministry of Science and Technology under the Scientific Schools grant 6501.2010.2.
\Appendix

\section{Commutation relations in the $S$-module} \label{comm_c}

Let $W^{(s)}_{\alpha}(z)$ be the highest weight state in the $S$-module, and $V^{(s)\pm}_{\alpha}(z)=G^{\pm}_{-1/3}W^{(s)}_{\alpha}(z)$ be its first fractional descendants. In the work \cite{Argyres:1993hz} one can find the OPEs for the operators $T(z)$ and $G^{\pm}(z)$ with the vertex operators in the $S$-module
\begin{align}\label{}
    & T(u)W^{(s)}_{\alpha}(z)=\frac{\Delta^{(s)}_{\alpha}}{(u-z)^2}W^{(s)}_{\alpha}(z)+\frac{1}{u-z} \partial W^{(s)}_{\alpha}(z)+\cdots, \label{opetw} \\
    & T(u)V^{(s)\pm}_{\alpha}(z)=\frac{\Delta^{(s)}_{\alpha}+\frac{1}{3}}{(u-z)^2}V^{(s)\pm}_{\alpha}(z)+\frac{1}{u-z} \partial V^{(s)\pm}_{\alpha}(z)+\cdots, \label{opetv} \\
    & G^{\pm}(u) W^{(s)}_{\alpha}(z)=\frac{1}{u-z} V^{(s)\pm}_{\alpha}(z)+\cdots, \label{opegw} \\
    & G^{\pm}(u) V^{(s)\pm}_{\alpha}(z)=\frac{\lambda^{\pm}}{2(u-z)^{\frac{4}{3}}} V^{(s)\mp}_{\alpha}(z)+\frac{1}{(u-z)^{\frac{1}{3}}}\left(\frac{\lambda^{\pm}}{3\Delta^{(s)}_\alpha+1}\partial V^{(s)\mp}_{\alpha}(z)+\tilde{V}^{(s)\mp}_{\alpha}(z) \right)+\cdots, \label{OPE_GV+} \\
    & G^{\pm}(u) V^{(s)\mp}_{\alpha}(z)=\frac{\Delta^{(s)}_{\alpha}}{(u-z)^{\frac{5}{3}}} W^{(s)}_{\alpha}(z)+\frac{1}{(u-z)^{\frac{2}{3}}}\left( \frac{1}{2} \partial W^{(s)}_{\alpha}(z) \pm \tilde{W}^{(s)}_{\alpha}(z) \right)\cdots, \label{OPE_GV-}
\end{align}
\noindent where the vertex operators $\tilde{W}^{(s)}_{\alpha}(z)$ and $\tilde{V}^{(s)\pm}(z)$ correspond to the states

\begin{align}\label{}
    & | \tilde{W}^{(s)}_{\alpha} \rangle=\left( G^{+}_{-\frac{2}{3}}G^{-}_{-\frac{1}{3}}-G^{-}_{-\frac{2}{3}}G^{+}_{-\frac{1}{3}} \right) | W^{(s)}_{\alpha} \rangle, \label{desc_1} \\
    & | \tilde{V}^{(s)\pm} \rangle=G^{\mp}_{-1} | V^{(s)\mp}_{\alpha} \rangle-\frac{\lambda^{\mp}}{3h_s+1}L_{-1} |V^{(s)\pm}_{\alpha} \rangle. \label{desc_2}
\end{align}
Then due to the fact that the OPEs (\ref{opetw}), (\ref{opetv}) and (\ref{opegw}) contain only integer powers it is easy to evaluate the corresponding commutation relations
\begin{align}\label{}
     [L_m, W^{(s)}_{\alpha}(z) ]&=\oint_z \frac{du}{2\pi i} u^{m+1} T(u) W^{(s)}_{\alpha}(z)= \oint_z \frac{du}{2\pi i} u^{m+1} \Bigl(\frac{\Delta^{(s)}_{\alpha}}{(u-z)^2}W^{(s)}_{\alpha}(z)+\frac{1}{u-z} \partial W^{(s)}_{\alpha}(z)+\cdots \Bigr)= \notag \\
    &  =z^{m}\partial W^{(s)}_{\alpha}(z)+(m+1)\Delta^{(s)}_{\alpha} W^{(s)}_{\alpha}(z), \\
     [L_m, V^{(s)\pm}_{\alpha}(z)] &=\oint_z \frac{du}{2\pi i} u^{m+1} T(u) V^{(s)\pm}_{\alpha}(z)= \oint_z \frac{du}{2\pi i} u^{m+1} \Bigl(\frac{\Delta^{(s)}_{\alpha}+\frac{1}{3}}{(u-z)^2}V^{(s)\pm}_{\alpha}(z)+\frac{1}{u-z} \partial V^{(s)\pm}_{\alpha}(z)+\cdots \Bigr)= \notag \\
    & =z^{m}\partial V^{(s)\pm}_{\alpha}(z)+(m+1)\Bigl(\Delta^{(s)}_{\alpha}+\frac{1}{3} \Bigr) V^{(s)\pm}_{\alpha}(z), \\
     [G^{\pm}_r, W^{(s)}_{\alpha}(z) ] &=\oint_z \frac{du}{2\pi i} u^{r+\frac{1}{3}} G^{\pm}(u) W^{(s)}_{\alpha}(z)=\oint_z \frac{du}{2\pi i} u^{r+\frac{1}{3}} \Bigl(\frac{V^{(s)\pm}_{\alpha}(z)}{u-z}+\ldots \Bigr)=z^{r+\frac{1}{3}}V^{(s)\pm}_{\alpha}(z).
\end{align}
Although due to the presence of the fractional powers in the OPEs (\ref{OPE_GV+}) and (\ref{OPE_GV-}) it is impossible to derive the commutation relations for these vertex operators, but it is possible to derive the generalized commutation relations in analogy with the modes of $G^{\pm}(u)$. To do this consider the integral with $p \in \mathbb{Z}$ with the points $z$ and $0$ inside the integration contour
\begin{align}\label{}
    \oint_{incl. z} \frac{du}{2\pi i} u^{r+\frac{1}{3}} (u-z)^{p+\frac{1}{3}} G^{\pm}(u) V^{(s)\pm}_{\alpha}(z)&=\sum_{l=0}^{+\infty} C^{\left(p+\frac{1}{3} \right)}_l z^l \oint \frac{du}{2\pi i} u^{r+p+\frac{2}{3}-l} G^{\pm}(u) V^{(s)\pm}_{\alpha}(z) = \notag \\
    & =  \sum_{l=0}^{+\infty} C^{\left(p+\frac{1}{3} \right)}_l z^l G^{\pm}_{r+p-l+\frac{1}{3}} V^{(s)\pm}_{\alpha}(z) =(\textrm{for}\; p=-1)  \notag\\
    &=\sum_{l=0}^{+\infty} C^{\left(-\frac{2}{3} \right)}_l z^l G^{\pm}_{r-l-\frac{2}{3}} V^{(s)\pm}_{\alpha}(z). \label{terml}
\end{align}
On the other side this integral can be divided into the integral over the point $z$, which is evaluated via the OPE (\ref{OPE_GV+}), and the integral around $0$ with $z$ outside this contour
\begin{align}\label{}
     \oint_{incl \; z} \frac{du}{2\pi i} u^{r+\frac{1}{3}} (u-z)^{p+\frac{1}{3}} G^{\pm}(u) V^{(s)\pm}_{\alpha}(z)=& \oint_z \frac{du}{2\pi i} u^{r+\frac{1}{3}} (u-z)^{p+\frac{1}{3}} G^{\pm}(u) V^{(s)\pm}_{\alpha}(z)+ \notag \\
    & +\oint_{excl \; z} \frac{du}{2\pi i} u^{r+\frac{1}{3}} (u-z)^{p+\frac{1}{3}} G^{\pm}(u) V^{(s)\pm}_{\alpha}(z).
\end{align}
We make the further calculations for $p=-1$. The first integral on the right hand side can be easily evaluated using the OPE (\ref{OPE_GV+})

\begin{align}\label{}
     \oint_z \frac{du}{2\pi i} u^{r+\frac{1}{3}} (u-z)^{-\frac{2}{3}} G^{\pm}(u) V^{(s)\pm}_{\alpha}(z)&=\oint_z \frac{du}{2\pi i} u^{r+\frac{1}{3}} \left( \frac{\lambda^{\pm}}{2(u-z)^{2}} V^{(s)\mp}_{\alpha}(z)+ \right. \notag \\
    & \left.\qquad +\frac{1}{u-z}\Bigl(\frac{\lambda^{\pm}}{3\Delta^{(s)}_\alpha+1}\partial V^{(s)\mp}_{\alpha}(z)+\tilde{V}^{(s)\mp}_{\alpha}(z) \Bigr)+\cdots \right)= \notag \\
    & =z^{r-\frac{2}{3}}\frac{\lambda^{\pm}}{2}\Bigl(r+\frac{1}{3}\Bigr) V^{(s)\mp}_{\alpha}(z)+z^{r+\frac{1}{3}} \Bigl( \frac{\lambda^{\pm}}{3h_s+1}\partial V^{(s)\mp}_{\alpha}(z)+\tilde{V}^{(s)\mp}_{\alpha}(z)\Bigr). \label{termope}
\end{align}
The second integral on the right hand side must be evaluated using the abelian braiding of $G^{\pm}$ and $V^{\pm}_s$, which gives the factor $e^{\frac{2i\pi}{3}}$.

\begin{align}\label{}
    \oint_{excl \; z} \frac{du}{2\pi i} u^{r+\frac{1}{3}} (u-z)^{p+\frac{1}{3}} G^{\pm}(u) V^{(s)\pm}_{\alpha}(z)&=e^{i \pi \left(p+\frac{1}{3} \right)} e^{\frac{2i \pi}{3}} \oint_{excl \; z} \frac{du}{2\pi i} u^{r+\frac{1}{3}} (z-u)^{p+\frac{1}{3}} V^{(s)\pm}_{\alpha}(z) G^{\pm}(u)= \notag \\
    & =(-1)^{p+1} \oint_{excl \; z} \frac{du}{2\pi i} u^{r+\frac{1}{3}} (z-u)^{p+\frac{1}{3}} V^{(s)\pm}_{\alpha}(z) G^{\pm}(u)= \notag \\
    & =(-1)^{p+1} \sum_{l=0}^{+\infty} C^{\left(p+\frac{1}{3} \right)}_l z^{p-l+\frac{1}{3}} \oint_{excl \; z} \frac{du}{2\pi i} u^{r+l+\frac{1}{3}} V^{(s)\pm}_{\alpha}(z) G^{\pm}(u)= \notag \\
    & = (-1)^{p+1} \sum_{l=0}^{+\infty} C^{\left(p+\frac{1}{3}\right)}_l z^{p-l+\frac{1}{3}} V^{(s)\pm}_{\alpha}(z) G^{\pm}_{r+l}=(\textrm{for}\; p=-1)\notag\\
    &=\sum_{l=0}^{+\infty} C^{\left(-\frac{2}{3}\right)}_l z^{-l-\frac{2}{3}} V^{(s)\pm}_{\alpha}(z) G^{\pm}_{r+l}. \label{termr}
\end{align}
Summarizing (\ref{terml}), (\ref{termope}) and (\ref{termr}) we are able to write the generalized commutation relations for $p=-1$
\begin{align}\label{}
    &\sum_{l=0}^{+\infty} C^{(-\frac{2}{3})}_l \left( z^{l} G^{\pm}_{r-l-\frac{2}{3}} V^{(s)\pm}_{\alpha}(z)-z^{-l-\frac{2}{3}} V^{(s)\pm}_{\alpha}(z) G^{\pm}_{r+l} \right)=\notag\\
    &\qquad\qquad\qquad\qquad=z^{r-\frac{2}{3}}\frac{\lambda^{\pm}}{2}\left(r+\frac{1}{3} \right) V^{(s)\mp}_{\alpha}(z)+z^{r+\frac{1}{3}} \left( \frac{\lambda^{\pm}}{3h_s+1}\partial V^{(s)\mp}_{\alpha}(z)+\tilde{V}^{(s)\mp}_{\alpha}(z) \right),
\end{align}
where the descendant $V^{(s)\pm}_{\alpha}(z)$ is given by the formula (\ref{desc_2}).

Then with the opposite $\mathbb{Z}_3$-charges
\begin{align}\label{}
     \oint_{incl \; z} \frac{du}{2\pi i} u^{r+\frac{1}{3}} (u-z)^{p+\frac{2}{3}} G^{\pm}(u) V^{(s)\mp}_{\alpha}(z)&=\sum_{l=0}^{+\infty} C^{\left(p+\frac{2}{3} \right)}_l z^l \oint_{incl \; z} \frac{du}{2\pi i} u^{r+p+1-l} G^{\pm}(u) V^{(s)\mp}_{\alpha}(z) = \notag \\
    & = \ \sum_{l=0}^{+\infty} C^{\left(p+\frac{2}{3} \right)}_l z^l G^{\pm}_{r+p-l+\frac{2}{3}} V^{(s)\mp}_{\alpha}(z) = (\textrm{for}\; p=-1) \notag\\
    &=\sum_{l=0}^{+\infty} C^{\left(-\frac{1}{3} \right)}_l z^l G^{\pm}_{r-l-\frac{1}{3}} V^{(s)\mp}_{\alpha}(z). \label{terml1}
\end{align}
On the other side this integral can be divided into the integral over the point $z$, which is evaluated via the OPE (\ref{OPE_GV-}), and the integral around $0$ with $z$ outside this contour
\begin{align}\label{}
    \oint_{incl \; z} \frac{du}{2\pi i} u^{r+\frac{1}{3}} (u-z)^{p+\frac{2}{3}} G^{\pm}(u) V^{(s)\mp}_{\alpha}(z)=& \oint_z \frac{du}{2\pi i} u^{r+\frac{1}{3}} (u-z)^{p+\frac{2}{3}} G^{\pm}(u) V^{(s)\mp}_{\alpha}(z)+ \notag \\
    & +\oint_{excl \; z} \frac{du}{2\pi i} u^{r+\frac{1}{3}} (u-z)^{p+\frac{2}{3}} G^{\pm}(u) V^{(s)\mp}_{\alpha}(z). \label{int_eval_2}
\end{align}
To evaluate the first term on the right hand side of (\ref{int_eval_2}) we use the OPE (\ref{frac_ope1}) (we again set $p=-1$)
\begin{align}\label{}
     \oint_z \frac{du}{2\pi i} u^{r+\frac{1}{3}} (u-z)^{-\frac{1}{3}} G^{\pm}(u) V^{(s)\mp}_{\alpha}(z)&=\oint_z \frac{du}{2\pi i} u^{r+\frac{1}{3}} \left( \frac{\Delta^{(s)}_{\alpha}}{(u-z)^{2}} W^{(s)}_{\alpha}(z)+ \right. \notag \\
    & \left.\qquad +\frac{1}{u-z} \left( \partial W^{(s)}_{\alpha}(z) \pm \tilde{W}^{(s)}_{\alpha}(z) \right)+\cdots \right)= \notag \\
    & =z^{r-\frac{2}{3}} \left(r+\frac{1}{3} \right) \Delta^{(s)}_{\alpha} W^{(s)}_{\alpha}(z)+z^{r+\frac{1}{3}} \left( \frac{1}{2}\partial W^{(s)}_{\alpha}(z) \pm \tilde{W}^{(s)}_{\alpha}(z) \right). \label{termope1}
\end{align}
The abelian braiding of the fields $G^{\pm}(u)$ and $V^{(s)\mp}(z)$ gives the factor $e^{-\frac{2i\pi}{3}}$
\begin{align}\label{}
    \oint_{excl \; z} \frac{du}{2\pi i} u^{r+\frac{1}{3}} (u-z)^{p+\frac{2}{3}} G^{\pm}(u) V^{(s)\mp}_{\alpha}(z)&=e^{i \pi \left(p+\frac{2}{3} \right)} e^{-\frac{2i \pi}{3}} \oint_{excl \; z} \frac{du}{2\pi i} u^{r+\frac{1}{3}} (z-u)^{p+\frac{2}{3}} V^{(s)\mp}_{\alpha}(z) G^{\pm}(u)= \notag \\
    & =(-1)^{p} \oint_{excl \; z} \frac{du}{2\pi i} u^{r+\frac{1}{3}} (z-u)^{p+\frac{2}{3}} V^{(s)\mp}_{\alpha}(z) G^{\pm}(u)= \notag \\
    & =(-1)^{p} \sum_{l=0}^{+\infty} C^{\left(p+\frac{2}{3} \right)}_l z^{p-l+\frac{2}{3}} \oint_{excl \; z} \frac{du}{2\pi i} u^{r+l+\frac{1}{3}} V^{(s)\mp}_{\alpha}(z) G^{\pm}(u)= \notag \\
    & = (-1)^{p} \sum_{l=0}^{+\infty} C^{\left(p+\frac{2}{3} \right)}_l z^{p-l+\frac{2}{3}} V^{(s)\mp}_{\alpha}(z) G^{\pm}_{r+l}= (\textrm{for}\; p=-1)\notag\\
    &=-\sum_{l=0}^{+\infty} C^{\left(-\frac{1}{3} \right)}_l z^{-l-\frac{1}{3}} V^{(s)\mp}_{\alpha}(z) G^{\pm}_{r+l}. \label{termr1}
\end{align}
After that summarizing (\ref{terml1}), (\ref{termope1}) and (\ref{termr1}) we are able to write the generalized commutation relations
\begin{align}\label{}
    &\sum_{l=0}^{+\infty} C^{\left(-\frac{1}{3} \right)}_l \left( z^{l} G^{\pm}_{r-l-\frac{1}{3}} V^{(s)\mp}_{\alpha}(z) + z^{-l-\frac{1}{3}} V^{(s)\mp}_{\alpha}(z) G^{\pm}_{r+l} \right)=\notag\\
    &\qquad\qquad\qquad =  z^{r-\frac{2}{3}}\left(r+\frac{1}{3} \right) \Delta^{(s)}_{\alpha} W^{(s)}_{\alpha}(z)+z^{r+\frac{1}{3}}\left(\frac{1}{2}\partial W^{(s)}_{\alpha}(z) \pm \tilde{W}^{(s)}_{\alpha}(z) \right),
\end{align}
where the descendant $\tilde{W}^{(s)}_{\alpha}(z)$ is given by the formula (\ref{desc_1}).

\section{Commutation relations in the $D$-module} \label{comm_d}
Let $W^{(d)\pm}_{\alpha}(z)$ be the vertex operators of the primary states in the $D$-module. To find the commutation relations we need the OPEs \cite{Argyres:1993hz} of the fields $W^{(d)\pm}_{\alpha}(z)$ and their first conformal descendants $V^{(d)(\pm)}_{\alpha}$
\begin{align}\label{}
    & T(u)W^{(d)\pm}_{\alpha}(z)=\frac{\Delta^{(d)}_{\alpha}}{(u-z)^2}W^{(d)\pm}_{\alpha}(z)+\frac{1}{u-z}\partial W^{(d)\pm}_{\alpha}(z)+\cdots ,\label{dOPE_1} \\
    & T(u)V^{(d)(\pm)}_{\alpha}(z)=\frac{\Delta^{(d)}_{\alpha}+\frac{2}{3}}{(u-z)^2}V^{(d)(\pm)}_{\alpha}(z)+\frac{1}{u-z} \partial V^{(d)(\pm)}_{\alpha}(z)+\cdots ,\label{dOPE_2}\\
    & G^{\pm}(u)W^{(d)\pm}_{\alpha}(z)=\frac{\Lambda^{\pm}}{(u-z)^{\frac{4}{3}}}W^{(d)\pm}_{\alpha}(z)+\frac{1}{(u-z)^{\frac{1}{3}}}
    \left(\frac{2\Lambda^{\pm}}{3\Delta^{(d)}_{\alpha}} \partial W^{(d)\pm}_{\alpha}(z)+\tilde{W}^{(d)\pm}_{\alpha}(z) \right)+\cdots, \label{dOPE_5}
\end{align}

\begin{align}
     G^{\pm}(u)W^{(d)\mp}_{\alpha}(z)=&\frac{1}{2(u-z)^{\frac{2}{3}}}\left( V^{(d)(+)}_{\alpha}(z) \pm V^{(d)(+)}_{\alpha}(z) \right)+\cdots, \label{dOPE_6} \\
     G^{\pm}(u)V^{(d)(+)}_{\alpha}(z)=&\frac{\Delta^{(d)}_{\alpha}
    +\frac{c}{12}+\lambda^{\pm}\Lambda^{\mp}}{(u-z)^2}W^{(d)\pm}_{\alpha}(z)+ \notag \\
    & +\frac{1}{u-z}\left( \frac{\Delta^{(d)}_{\alpha}+\frac{c}{12}+\lambda^{\pm}\Lambda^{\mp}}{3\Delta^{(s)}_{\alpha}} \partial W^{(d)\pm}_{\alpha}(z)-\left(\Lambda^{\pm}-\frac{1}{2}\lambda^{\pm} \right)\tilde{W}^{(d)\pm}_{\alpha}(z) \right)+\cdots. \label{dOPE_3}
\end{align}
\begin{align}
     G^{\pm}(u)V^{(d)(-)}_{\alpha}(z)=&\mp\frac{\Delta^{(d)}_{\alpha}+\frac{c}{12}-\lambda^{\pm}\Lambda^{\mp}}{(u-z)^2}W^{(d)\pm}_{\alpha}(z)+ \notag \\
    & +\frac{1}{u-z}\left( \mp\frac{\Delta^{(d)}_{\alpha}+\frac{c}{12}-\lambda^{\pm}\Lambda^{\mp}}{3\Delta^{(s)}_{\alpha}} \partial W^{(d)\pm}_{\alpha}(z) \pm \left(\Lambda^{\pm}+\frac{1}{2}\lambda^{\pm} \right)\tilde{W}^{(d)\pm}_{\alpha}(z) \right)+\cdots, \label{dOPE_4}
\end{align}
\noindent where the vertex operators $V^{(d)(\pm)}_{\alpha}(z)$ and $\tilde{W}^{(d)\pm}_{\alpha}(z)$ correspond to the states
\begin{align}\label{}
    & | V^{(d)(\pm)}_{\alpha} \rangle=G^{+}_{-\frac{2}{3}} | W^{(d)-}_{\alpha} \rangle \pm G^{-}_{-\frac{2}{3}} | W^{(d)+}_{\alpha} \rangle, \label{descd_1} \\
    & | \tilde{W}^{(d)\pm}_{\alpha} \rangle=G^{\mp}_{-1} | W^{(d)\mp}_{\alpha} \rangle-\frac{2\Lambda^{\mp}}{3\Delta^{(d)}_{\alpha}} L_{-1} | W^{(d)\pm}_{\alpha} \rangle. \label{descd_2}
\end{align}

Due to the fact that the OPEs (\ref{dOPE_1}), (\ref{dOPE_2}), (\ref{dOPE_3}) and (\ref{dOPE_4}) contain only integer powers of $u-z$, we are able to write the commutation relations

\begin{align}\label{}
     [L_m, W^{(d)\pm}_{\alpha}(z) ]&=\oint_z \frac{du}{2\pi i} u^{m+1} T(u) W^{(d)\pm}_{\alpha}(z)= \oint_z \frac{du}{2\pi i} u^{m+1} \Bigl(\frac{\Delta^{(d)}_{\alpha}}{(u-z)^2}W^{(d)\pm}_{\alpha}(z)+\frac{1}{u-z} \partial W^{(d)\pm}_{\alpha}(z)+\cdots  \Bigr)= \notag \\
    & =z^{m}\partial W^{(d)\pm}_{\alpha}(z)+(m+1)\Delta^{(d)}_{\alpha} W^{(d)\pm}_{\alpha}(z)  \\
    [L_m, V^{(d)(\pm)}_{\alpha}(z) ]&=\oint_z \frac{du}{2\pi i} u^{m+1} T(u) V^{(d)(\pm)}_{\alpha}(z)= \oint_z \frac{du}{2\pi i} u^{m+1}  \Bigl(\frac{\Delta^{(d)}_{\alpha}+\frac{2}{3}}{(u-z)^2}V^{(s)(\pm)}_{\alpha}(z)+\frac{1}{u-z} \partial V^{(d)(\pm)}_{\alpha}(z)+..\Bigr)= \notag \\
    & =z^{m}\partial V^{(d)(\pm)}_{\alpha}(z)+(m+1)\left(\Delta^{(d)}_{\alpha}+\frac{2}{3} \right) V^{(d)(\pm)}_{\alpha}(z),
  \end{align}
  \begin{align}
     [G^{\pm}_r, V^{(d)+}_{\alpha}(z) ]&=\oint_z \frac{du}{2\pi i} u^{r+\frac{1}{3}} G^{\pm}(u) V^{(d)+}_{\alpha}(z)=\oint_z \frac{du}{2\pi i} u^{r+\frac{1}{3}} \left(\frac{\Delta^{(d)}_{\alpha}+\frac{c}{12}+\lambda^{\pm}\Lambda^{\mp}}{(u-z)^2}W^{(d)\pm}_{\alpha}(z)+ \right. \notag \\
    & \left. \qquad +\frac{1}{u-z} \Bigl( \frac{\Delta^{(d)}_{\alpha}+\frac{c}{12}+\lambda^{\pm}\Lambda^{\mp}}{3\Delta^{(s)}_{\alpha}} \partial W^{(d)\pm}_{\alpha}(z)- \Bigl(\Lambda^{\pm}-\frac{1}{2}\lambda^{\pm} \Bigr)\tilde{W}^{(d)\pm}_{\alpha}(z)  \Bigr)+\cdots \right)= \notag \\
    & =z^{r-\frac{2}{3}}\left(r+\frac{1}{3} \right) \left(\Delta^{(d)}_{\alpha}+\frac{c}{12}+\lambda^{\pm} \Lambda^{\mp} \right) W^{(d)\pm}_{\alpha}(z)+z^{r+\frac{1}{3}}\frac{\Delta^{(d)}_{\alpha}+\frac{c}{12}+\lambda^{\pm} \Lambda^{\mp}}{3\Delta^{(d)}_{\alpha}} \partial W^{(d)\pm}_{\alpha}(z)-\notag\\
    &\qquad -z^{r+\frac{1}{3}} \left(\Lambda^{\pm}-\frac{1}{2}\lambda^{\pm} \right) \tilde{W}^{(d)\pm}_{\alpha}(z),
      \end{align}
     \begin{align}
     [G^{\pm}_r, V^{(d)-}_{\alpha}(z) ]&=\oint_z \frac{du}{2\pi i} u^{r+\frac{1}{3}} G^{\pm}(u) V^{(d)-}_{\alpha}(z)=\oint_z \frac{du}{2\pi i} u^{r+\frac{1}{3}} \left(\mp\frac{\Delta^{(d)}_{\alpha}+\frac{c}{12}-\lambda^{\pm}\Lambda^{\mp}}{(u-z)^2}W^{(d)\pm}_{\alpha}(z)+ \right. \notag \\
    & \left. \qquad +\frac{1}{u-z} \Bigl( \mp\frac{\Delta^{(d)}_{\alpha}+\frac{c}{12}-\lambda^{\pm}\Lambda^{\mp}}{3\Delta^{(s)}_{\alpha}} \partial W^{(d)\pm}_{\alpha}(z)\pm \Bigl(\Lambda^{\pm}+\frac{1}{2}\lambda^{\pm}  \Bigr)\tilde{W}^{(d)\pm}_{\alpha}(z) \Bigr)+\cdots \right)= \notag \\
    & \qquad \mp z^{r-\frac{2}{3}}\left(r+\frac{1}{3} \right) \left(\Delta^{(d)}_{\alpha}+\frac{c}{12}-\lambda^{\pm} \Lambda^{\mp} \right) W^{(d)\pm}_{\alpha}(z) \mp z^{r+\frac{1}{3}}\frac{\Delta^{(d)}_{\alpha}+\frac{c}{12}-\lambda^{\pm} \Lambda^{\mp}}{3\Delta^{(d)}_{\alpha}} \partial W^{(d)\pm}_{\alpha}(z) \pm \notag\\
    &\qquad \pm z^{r+\frac{1}{3}} \left(\Lambda^{\pm}+\frac{1}{2}\lambda^{\pm} \right) \tilde{W}^{(d)\pm}_{\alpha}(z),
\end{align}

\noindent where the descendant $\tilde{W}^{(d)\pm}_{\alpha}(z)$ is given by the formula (\ref{descd_2}).

Analogously to the case of the $S$-module
\begin{align}\label{}
    \oint_{incl \; z} \frac{du}{2\pi i} u^{r+\frac{1}{3}} (u-z)^{p+\frac{1}{3}} G^{\pm}(u) W^{(d)\pm}_{\alpha}(z)=& \oint_z \frac{du}{2\pi i} u^{r+\frac{1}{3}} (u-z)^{p+\frac{1}{3}} G^{\pm}(u) W^{(d)\pm}_{\alpha}(z)+ \notag \\
    & +\oint_{excl \; z} \frac{du}{2\pi i} u^{r+\frac{1}{3}} (u-z)^{p+\frac{1}{3}} G^{\pm}(u) W^{(d)\pm}_{\alpha}(z).
\end{align}

The contours encircling the origin can be easily evaluated using the abelian braiding of $G^{\pm}(u)$ with $W^{(d)\pm}_{\alpha}(z)$, which gives in this case the factor $e^{\frac{2i\pi}{3}}$

\begin{align}\label{}
    & \oint_{incl \; z} \frac{du}{2\pi i} u^{r+\frac{1}{3}} (u-z)^{p+\frac{1}{3}} G^{\pm}(u) W^{(d)\pm}_{\alpha}(z)=\sum_{l=0}^{+\infty} C^{(p+\frac{1}{3})}_l z^l G^{\pm}_{r+p-l+\frac{1}{3}} W^{(d)\pm}_{\alpha}(z), \label{incl} \\
    & \oint_{excl \; z} \frac{du}{2\pi i} u^{r+\frac{1}{3}} (u-z)^{p+\frac{1}{3}} G^{\pm}(u) W^{(d)\pm}_{\alpha}(z)=(-1)^{p+1} \sum_{l=0}^{+\infty} C^{(p+\frac{1}{3})}_l z^{p-l+\frac{1}{3}} W^{(d)\pm}_{\alpha}(z) G^{\pm}_{r+l}. \label{excl}
\end{align}

The term with the contour encircling the point $z$ is calculated using the OPE (\ref{dOPE_5}) in the $D$-module (we again set $p=-1$)
\begin{align}\label{}
     \oint_z \frac{du}{2\pi i} u^{r+\frac{1}{3}} (u-z)^{-\frac{2}{3}} G^{\pm}(u) W^{(d)\pm}_{\alpha}(z)&=\oint_z \frac{du}{2\pi i} u^{r+\frac{1}{3}} \left( \frac{\Lambda^{\pm}}{(u-z)^{2}}W^{(d)\pm}_{\alpha}(z)+ \right. \notag \\
    & \left. \qquad +\frac{1}{u-z}
    \left(\frac{2\Lambda^{\pm}}{3\Delta^{(d)}_{\alpha}} \partial W^{(d)\pm}_{\alpha}(z)+\tilde{W}^{(d)\pm}_{\alpha}(z) \right)+\cdots \right)= \notag \\
    & =z^{r-\frac{2}{3}}\left(r+\frac{1}{3} \right)\Lambda^{\pm} W^{(d)\mp}_{\alpha}(z)+z^{r+\frac{1}{3}}\left( \frac{2\Lambda^{\pm}}{3\Delta^{(d)}_{\alpha}}\partial W^{(d)\mp}_{\alpha}(z)+\tilde{W}^{(d)\mp}_{\alpha}(z) \right). \label{termope2}
\end{align}
Then summarizing the derived equations (\ref{incl}), (\ref{excl}) and (\ref{termope2}) we have
\begin{align}\label{}
    & \sum_{l=0}^{+\infty} C^{(-\frac{2}{3})}_l \left( z^{l} G^{\pm}_{r-l-\frac{2}{3}} W^{(d)\pm}_{\alpha}(z)-z^{-l-\frac{2}{3}} W^{(d)\pm}_{\alpha}(z) G^{\pm}_{r+l} \right)= \notag \\
    &\qquad\qquad\qquad\qquad =z^{r-\frac{2}{3}}\left(r+\frac{1}{3} \right)\Lambda^{\pm} W^{(d)\mp}_{\alpha}(z)+z^{r+\frac{1}{3}}\left( \frac{2\Lambda^{\pm}}{3\Delta^{(d)}_{\alpha}}\partial W^{(d)\mp}_{\alpha}(z)+\tilde{W}^{(d)\mp}_{\alpha}(z) \right).
\end{align}
The same thing with the opposite signs of $\mathbb{Z}_3$-charge
\begin{align}\label{}
     \oint_{incl \; z} \frac{du}{2\pi i} u^{r+\frac{1}{3}} (u-z)^{p+\frac{2}{3}} G^{\pm}(u) W^{(d)\mp}_{\alpha}(z)=& \oint_z \frac{du}{2\pi i} u^{r+\frac{1}{3}} (u-z)^{p+\frac{2}{3}} G^{\pm}(u) W^{(d)\mp}_{\alpha}(z)+ \notag \\
    & +\oint_{excl \; z} \frac{du}{2\pi i} u^{r+\frac{1}{3}} (u-z)^{p+\frac{2}{3}} G^{\pm}(u) W^{(d)\mp}_{\alpha}(z).
\end{align}
The contours encircling the origin can be easily evaluated using the abelian braiding of $G^{\pm}(u)$ with $W^{(d)\mp}_{\alpha}(z)$, which gives in this case the factor $e^{-\frac{2i\pi}{3}}$
\begin{align}\label{}
    & \oint_{incl \; z} \frac{du}{2\pi i} u^{r+\frac{1}{3}} (u-z)^{p+\frac{2}{3}} G^{\pm}(u) W^{\mp}_d(z)=\sum_{l=0}^{+\infty} C^{(p+\frac{2}{3})}_l z^l G^{\pm}_{r+p-l+\frac{2}{3}} W^{\pm}_d(z), \label{incl1} \\
    & \oint_{excl \; z} \frac{du}{2\pi i} u^{r+\frac{1}{3}} (u-z)^{p+\frac{2}{3}} G^{\pm}(u) W^{\mp}_d(z)=(-1)^{p} \sum_{l=0}^{+\infty} C^{(p+\frac{2}{3})}_l z^{p-l+\frac{2}{3}} W^{\mp}_d(z) G^{\pm}_{r+l}. \label{excl1}
\end{align}
The term with the contour encircling the point $z$ is calculated using the OPE (\ref{dOPE_6}) in the $D$-module (we again set $p=-1$)
\begin{align}\label{}
     \oint_z \frac{du}{2\pi i} u^{r+\frac{1}{3}} (u-z)^{-\frac{1}{3}} G^{\pm}(u) W^{(d)\mp}_{\alpha}(z)&=\oint_z \frac{du}{2\pi i} u^{r+\frac{1}{3}} \left( \frac{1}{2(u-z)}\left( V^{(d)(+)}_{\alpha}(z) \pm V^{(d)(+)}_{\alpha}(z) \right)+\cdots \right)= \notag \\
    & =\frac{1}{2}z^{r+\frac{1}{3}} \left(V^{(d)(+)}_{\alpha}(z) \pm V^{(d)(-)}_{\alpha}(z) \right). \label{termope3}
\end{align}
After the above operations, summarizing (\ref{incl1}), (\ref{excl1}) and (\ref{termope3}), we are able to write the generalized commutation relations
\begin{equation}\label{}
    \sum_{l=0}^{+\infty} C^{(-\frac{1}{3})}_l \left( z^l G^{\pm}_{r-l-\frac{1}{3}} W^{(d)\mp}_{\alpha}(z) + z^{-l-\frac{1}{3}} W^{(d)\mp}_{\alpha}(z) G^{\pm}_{r+l} \right)=\frac{1}{2}z^{r+\frac{1}{3}} \left(V^{(d)(+)}_{\alpha}(z) \pm V^{(d)(-)}_{\alpha}(z) \right),
\end{equation}

\noindent where the descendants $V^{(d)(\pm)}_{\alpha}(z)$ is given by the formula (\ref{descd_1}).

\section{Gram/Shapovalov matrix and matrix elements at the level 7/4} \label{app_c}
The Gram/Shapovalov matrix at the level 1
\small
\begin{equation}\label{}
    \left(
      \begin{array}{cccc}
        -\left(\frac{2}{3}\Delta^{(d)}_{\alpha}+\frac{5c}{9} \right) & \frac{4}{3}\sqrt{\frac{c-8}{6}} \sqrt{\frac{c}{24}-\Delta^{(d)}_{\alpha}} & -2\left(\Delta^{(d)}_{\alpha}+\frac{c}{12} \right) & 2\sqrt{\frac{c-8}{6}} \sqrt{\frac{c}{24}-\Delta^{(d)}_{\alpha}} \\
        \frac{4}{3}\sqrt{\frac{c-8}{6}} \sqrt{\frac{c}{24}-\Delta^{(d)}_{\alpha}} & -\left(\frac{2}{3}\Delta^{(d)}_{\alpha}+\frac{5c}{9} \right) & 2\sqrt{\frac{c-8}{6}} \sqrt{\frac{c}{24}-\Delta^{(d)}_{\alpha}} & -2\left(\Delta^{(d)}_{\alpha}+\frac{c}{12} \right) \\
        -2\left(\Delta^{(d)}_{\alpha}+\frac{c}{12} \right) & 2\sqrt{\frac{c-8}{6}} \sqrt{\frac{c}{24}-\Delta^{(d)}_{\alpha}} & -2\left(\Delta^{(d)}_{\alpha}+\frac{2}{3} \right)\left(\Delta^{(d)}_{\alpha}+\frac{c}{12} \right) & 2\left(\Delta^{(d)}_{\alpha}+\frac{2}{3} \right)\sqrt{\frac{c-8}{6}} \sqrt{\frac{c}{24}-\Delta^{(d)}_{\alpha}} \\
        2\sqrt{\frac{c-8}{6}} \sqrt{\frac{c}{24}-\Delta^{(d)}_{\alpha}} & -2\left(\Delta^{(d)}_{\alpha}+\frac{c}{12} \right) & 2\left(\Delta^{(d)}_{\alpha}+\frac{2}{3} \right)\sqrt{\frac{c-8}{6}} \sqrt{\frac{c}{24}-\Delta^{(d)}_{\alpha}} & -2\left(\Delta^{(d)}_{\alpha}+\frac{2}{3} \right)\left(\Delta^{(d)}_{\alpha}+\frac{c}{12} \right) \\
      \end{array}
    \right).
\end{equation}
\normalsize
The corresponding matrix elements
\begin{align}\label{}
    & \left\langle m_1;0 \left| W_{\alpha_1}(1) \right| 1 \right\rangle_{7/4}=-\left(\mathbb{C}^{(+)\alpha}_{m_1, \alpha_1}+\mathbb{C}^{(-)\alpha}_{m_1, \alpha_1}\right), \\
    & \left\langle m_1;0 \left| W_{\alpha_1}(1) \right| 2 \right\rangle_{7/4}=-\left(\mathbb{C}^{(+)\alpha}_{m_1, \alpha_1}-\mathbb{C}^{(-)\alpha}_{m_1, \alpha_1}\right), \notag\\
    & \left\langle m_1;0 \left| W_{\alpha_1}(1) \right| 3 \right\rangle_{7/4}=\left( \Delta^{(s)}_{m_1}-\Delta^{(s)}_{\alpha_1}-\Delta^{(d)}_{\alpha}-\frac{2}{3} \right) \times \left(\mathbb{C}^{(+)\alpha}_{m_1, \alpha_1}+\mathbb{C}^{(-)\alpha}_{m_1, \alpha_1}\right), \notag\\
    & \left\langle m_1;0 \left| W_{\alpha_1}(1) \right| 4 \right\rangle_{7/4}=\left( \Delta^{(s)}_{m_1}-\Delta^{(s)}_{\alpha_1}-\Delta^{(d)}_{\alpha}-\frac{2}{3} \right) \times \left(\mathbb{C}^{(+)\alpha}_{m_1, \alpha_1}-\mathbb{C}^{(-)\alpha}_{m_1, \alpha_1}\right), \notag\\
    & {}_{7/4}\langle 1| W_{\alpha_2}(z) | m_2;0 \rangle=-z^{\Delta^{(d)}_{\alpha}+\frac{5}{3}-\Delta^{(s)}_{\alpha_2}-\Delta^{(s)}_{m_2}} \times \left(\mathbb{C}^{(+)m_2}_{\alpha, \alpha_2}+\mathbb{C}^{(-)m_2}_{\alpha, \alpha_2}\right), \notag\\
    & {}_{7/4}\langle 2| W_{\alpha_2}(z)| m_2;0 \rangle=-z^{\Delta^{(d)}_{\alpha}+\frac{5}{3}-\Delta^{(s)}_{\alpha_2}-\Delta^{(s)}_{m_2}} \times \left(\mathbb{C}^{(+)m_2}_{\alpha, \alpha_2}-\mathbb{C}^{(-)m_2}_{\alpha, \alpha_2}\right),\notag \\
    & {}_{7/4}\langle 3| W_{\alpha_2}(z) | m_2;0 \rangle=-\left( \Delta^{(d)}_{\alpha}+\frac{2}{3}+\Delta^{(s)}_{\alpha_2}-\Delta^{(s)}_{m_2} \right) z^{\Delta^{(d)}_{\alpha}+\frac{5}{3}-\Delta^{(s)}_{\alpha_2}-\Delta^{(s)}_{m_2}}\times \left(\mathbb{C}^{(+)m_2}_{\alpha, \alpha_2}+\mathbb{C}^{(-)m_2}_{\alpha, \alpha_2}\right),\notag\\
    & {}_{7/4}\langle 4| W_{\alpha_2}(z) | m_2;0 \rangle=-\left( \Delta^{(d)}_{\alpha}+\frac{2}{3}+\Delta^{(s)}_{\alpha_2}-\Delta^{(s)}_{m_2} \right) z^{\Delta^{(d)}_{\alpha}+\frac{5}{3}-\Delta^{(s)}_{\alpha_2}-\Delta^{(s)}_{m_2}}\times \left(\mathbb{C}^{(+)m_2}_{\alpha, \alpha_2}-\mathbb{C}^{(-)m_2}_{\alpha, \alpha_2}\right). \notag
\end{align}

\section{Gram/Shapovalov matrix and matrix elements at the level 2} \label{app_d}
The corresponding Gram/Shapovalov matrix is
\scriptsize
\begin{equation}\label{}
{ \tiny
    \left(
      \begin{array}{ccccc}
        \frac{2}{3}\Delta^{(s)}_{\alpha} \left(\Delta^{(s)}_{\alpha}{+}1 \right) \left(2\Delta^{(s)}_{\alpha}{+}\frac{c}{4}{+}1 \right){+}\left(\Delta^{(s)}_{\alpha}\right)^2 & {-}\frac{2}{3}\Delta^{(s)}_{\alpha} \left(\Delta^{(s)}_{\alpha}{+}1 \right) \left(2\Delta^{(s)}_{\alpha}{+}\frac{c}{4}{-}2 \right){+}\left(\Delta^{(s)}_{\alpha}\right)^2 & \frac{2}{3}\Delta^{(s)}_{\alpha} \left(2\Delta^{(s)}_{\alpha}{+}\frac{c}{4}{+}1 \right) & -\frac{2}{3}\Delta^{(s)}_{\alpha} \left(2\Delta^{(s)}_{\alpha}{+}\frac{c}{4}{-}2 \right) & 3\Delta^{(s)}_{\alpha} \\
        {-}\frac{2}{3}\Delta^{(s)}_{\alpha} \left(\Delta^{(s)}_{\alpha}{+}1 \right) \left(2\Delta^{(s)}_{\alpha}{+}\frac{c}{4}{-}2 \right){+}\left(\Delta^{(s)}_{\alpha}\right)^2 & \frac{2}{3}\Delta^{(s)}_{\alpha} \left(\Delta^{(s)}_{\alpha}{+}1 \right) \left(2\Delta^{(s)}_{\alpha}{+}\frac{c}{4}{+}1 \right){+}\left(\Delta^{(s)}_{\alpha}\right)^2 & -\frac{2}{3}\Delta^{(s)}_{\alpha} \left(2\Delta^{(s)}_{\alpha}{+}\frac{c}{4}{-}2 \right) & \frac{2}{3}\Delta^{(s)}_{\alpha} \left(2\Delta^{(s)}_{\alpha}{+}\frac{c}{4}{+}1 \right) & 3\Delta^{(s)}_{\alpha} \\
        \frac{2}{3}\Delta^{(s)}_{\alpha} \left(2\Delta^{(s)}_{\alpha}{+}\frac{c}{4}{+}1 \right) & -\frac{2}{3}\Delta^{(s)}_{\alpha} \left(2\Delta^{(s)}_{\alpha}{+}\frac{c}{4}{-}2 \right) & \frac{\Delta^{(s)}_{\alpha}}{9}\left(5\Delta^{(s)}_{\alpha}{+}5c{+}2 \right) & -\frac{\Delta^{(s)}_{\alpha}}{9} \left(\Delta^{(s)}_{\alpha}{+}c{-}8 \right) & \frac{7}{3}\Delta^{(s)}_{\alpha} \\
        -\frac{2}{3}\Delta^{(s)}_{\alpha} \left(2\Delta^{(s)}_{\alpha}{+}\frac{c}{4}{-}2 \right) & \frac{2}{3}\Delta^{(s)}_{\alpha} \left(2\Delta^{(s)}_{\alpha}{+}\frac{c}{4}{+}1 \right) & -\frac{\Delta^{(s)}_{\alpha}}{9} \left(\Delta^{(s)}_{\alpha}{+}c{-}8 \right) & \frac{\Delta^{(s)}_{\alpha}}{9}\left(5\Delta^{(s)}_{\alpha}{+}5c{+}2 \right) & \frac{7}{3}\Delta^{(s)}_{\alpha} \\
        3\Delta^{(s)}_{\alpha} & 3\Delta^{(s)}_{\alpha} & \frac{7}{3}\Delta^{(s)}_{\alpha} & \frac{7}{3}\Delta^{(s)}_{\alpha} & 4\left(\Delta^{(s)}_{\alpha}{+}\frac{c}{8} \right) \\
      \end{array}
    \right)
}
\end{equation}
\normalsize
The matrix elements have the corresponding form
\begin{align}\label{}
     \left\langle m_1;0 \left| W_{\alpha_1}(1) \right| 1 \right\rangle_{2}=&\frac{1}{2}\left(\Delta^{(s)}_{m_1}-\Delta^{(s)}_{\alpha_1}-\Delta^{(s)}_{\alpha} \right)\left(\Delta^{(s)}_{m_1}-\Delta^{(s)}_{\alpha_1}-\Delta^{(s)}_{\alpha}-1 \right) \times C^{\alpha}_{m_1,\alpha_1}+ \\
    &\quad +\left(\Delta^{(s)}_{m_1}-\Delta^{(s)}_{\alpha_1}-\Delta^{(s)}_{\alpha}-1 \right) \times \tilde{C}^{\alpha}_{m_1,\alpha_1},  \notag\\
     \left\langle m_1;0 \left| W_{\alpha_1}(1) \right| 2 \right\rangle_{2}=&\frac{1}{2}\left(\Delta^{(s)}_{m_1}-\Delta^{(s)}_{\alpha_1}-\Delta^{(s)}_{\alpha} \right)\left(\Delta^{(s)}_{m_1}-\Delta^{(s)}_{\alpha_1}-\Delta^{(s)}_{\alpha}-1 \right) \times C^{\alpha}_{m_1,\alpha_1}- \notag  \\
    &\quad -\left(\Delta^{(s)}_{m_1}-\Delta^{(s)}_{\alpha_1}-\Delta^{(s)}_{\alpha}-1 \right) \times \tilde{C}^{\alpha}_{m_1,\alpha_1}, \notag\\
     \left\langle m_1;0 \left| W_{\alpha_1}(1) \right| 3 \right\rangle_{2}=&-\frac{1}{2}\left(\Delta^{(s)}_{m_1}-\Delta^{(s)}_{\alpha_1}-\Delta^{(s)}_{\alpha} \right)\times C^{\alpha}_{m_1,\alpha_1}-\tilde{C}^{\alpha}_{m_1,\alpha_1}, \notag  \\
     \left\langle m_1;0 \left| W_{\alpha_1}(1) \right| 4 \right\rangle_{2}=&-\frac{1}{2}\left(\Delta^{(s)}_{m_1}-\Delta^{(s)}_{\alpha_1}-\Delta^{(s)}_{\alpha} \right)\times C^{\alpha}_{m_1,\alpha_1}-\tilde{C}^{\alpha}_{m_1,\alpha_1}, \notag \\
     \left\langle m_1;0 \left| W_{\alpha_1}(1) \right| 5 \right\rangle_{2}=&\left(2\Delta^{(s)}_{\alpha_1}+\Delta^{(s)}_{\alpha}-\Delta^{(s)}_{m_1} \right) \times C^{\alpha}_{m_1,\alpha_1}, \notag
\end{align}

\begin{align*}
    {}_{2}\langle 1| W_{\alpha_2}(z)| m_2;0 \rangle=&z^{\Delta^{(s)}_{\alpha}-\Delta^{(s)}_{\alpha_2}-\Delta^{(s)}_{m_2}+2} \left( \frac{1}{2}\left(\Delta^{(s)}_{\alpha}+1+\Delta^{(s)}_{\alpha_2}-\Delta^{(s)}_{m_2} \right)\left(\Delta^{(s)}_{\alpha}+\Delta^{(s)}_{\alpha_2}-\Delta^{(s)}_{m_2} \right)\times C^{m_2}_{\alpha,\alpha_2}- \right. \notag \\
    & \left.\qquad\qquad\qquad\qquad-\left(\Delta^{(s)}_{\alpha}+1+\Delta^{(s)}_{\alpha_2}-\Delta^{(s)}_{m_2} \right) \times \tilde{C}^{m_2}_{\alpha,\alpha_2} \right), \notag\\
     {}_{2}\langle 2| W_{\alpha_2}(z) | m_2;0 \rangle=&z^{\Delta^{(s)}_{\alpha}-\Delta^{(s)}_{\alpha_2}-\Delta^{(s)}_{m_2}+2} \left( \frac{1}{2}\left(\Delta^{(s)}_{\alpha}+1+\Delta^{(s)}_{\alpha_2}-\Delta^{(s)}_{m_2} \right)\left(\Delta^{(s)}_{\alpha}+\Delta^{(s)}_{\alpha_2}-\Delta^{(s)}_{m_2} \right)\times C^{m_2}_{\alpha,\alpha_2}+ \right. \notag \\
    & \left.\qquad\qquad\qquad\qquad+\left(\Delta^{(s)}_{\alpha}+1+\Delta^{(s)}_{\alpha_2}-\Delta^{(s)}_{m_2} \right) \times \tilde{C}^{m_2}_{\alpha,\alpha_2} \right), \notag\\
    {}_{2}\langle 3| W_{\alpha_2}(z) | m_2;0 \rangle=&z^{\Delta^{(s)}_{\alpha}-\Delta^{(s)}_{\alpha_2}-\Delta^{(s)}_{m_2}+2} \left( \frac{1}{2}\left(\Delta^{(s)}_{\alpha}+\Delta^{(s)}_{\alpha_2}-\Delta^{(s)}_{m_2} \right) \times C^{m_2}_{\alpha,\alpha_2}-\tilde{C}^{m_2}_{\alpha,\alpha_2} \right), \notag\\
     {}_{2}\langle 4| W_{\alpha_2}(z)| m_2;0 \rangle=&z^{\Delta^{(s)}_{\alpha}-\Delta^{(s)}_{\alpha_2}-\Delta^{(s)}_{m_2}+2} \left( \frac{1}{2}\left(\Delta^{(s)}_{\alpha}+\Delta^{(s)}_{\alpha_2}-\Delta^{(s)}_{m_2} \right)\times  C^{m_2}_{\alpha,\alpha_2}+\tilde{C}^{m_2}_{\alpha,\alpha_2} \right), \notag \\
     {}_{2}\langle 5| W_{\alpha_2}(z) | m_2;0 \rangle=&z^{\Delta^{(s)}_{\alpha}-\Delta^{(s)}_{\alpha_2}-\Delta^{(s)}_{m_2}+2} \left( 2\Delta^{(s)}_{\alpha_2}+\Delta^{(s)}_{\alpha}-\Delta^{(s)}_{m_2} \right) \times C^{m_2}_{\alpha,\alpha_2}.
\end{align*}


\bibliographystyle{MyStyle}
\bibliography{MyBib}

\end{document}